\documentclass[a4paper,12pt]{article}
\pdfoutput=1 % if your are submitting a pdflatex (i.e. if you have
\usepackage{jheppub_X} % for details on the use of the package, please
\usepackage[T1]{fontenc} % if needed
\usepackage{amsmath}
\usepackage{dsfont}
\usepackage{slashed}
\usepackage{bm}
\usepackage{enumerate}

\usepackage{longtable}
\usepackage{multirow}

\usepackage{hhline}
\usepackage{rotating}
\usepackage{array}
\usepackage{float}
\usepackage{afterpage}
\usepackage{anyfontsize}
\usepackage{enumitem}
\usepackage[dvipsnames,table]{xcolor}
\usepackage{arydshln}
\usepackage{blkarray}
\usepackage{physics}
\usepackage{cleveref}
\usepackage{mathtools}
\usepackage{xspace}
\usepackage{setspace}
\usepackage[normalem]{ulem}

\usepackage[most]{tcolorbox}
\definecolor{light-gray}{gray}{0.9}
\usepackage{bbold}

\makeatletter
\g@addto@macro\bfseries{\boldmath}
\makeatother

\crefname{table}{Table}{Tables}
\crefname{equation}{Eq.}{Eqs.}
\crefname{appendix}{App.}{Apps.}
\crefname{section}{Sec.}{Secs.}
\crefname{figure}{Fig.}{Figs.}

  {}% \end{boldenv}

\def\L{{\cal L}}
\def\O{{\cal O}}
\def\Dcal{{\cal D}}
\def\Acal{{\cal A}}
\def\Jcal{{\cal J}}

\def\brst{\text{\tiny BRST}}

\def\ie{\emph{i.e.}}
\def\eg{\emph{e.g.}}
\def\cf{\emph{cf.}}

\def\Pb#1{\widehat{P}_\beta^{#1}}

\def\bna{\beta_\text{\tiny NA}}

\def\Psl{\slashed{P}}
\def\qsl{\slashed{q}}
\def\ksl{\slashed{k}}
\def\Gsl{\slashed{G}}
\def\Rsl{\slashed{R}}
\def\Lsl{\slashed{L}}
\def\pdsl{\slashed\partial}
\def\seq{\simeq}
\def\eps{\varepsilon}

\def\ttr{{\tr_x}}
\newcommand{\s}{\hspace{0.8pt}}

\allowdisplaybreaks

\preprint{
\vspace{-8pt}
\begin{flushright}CERN-TH-2023-001\end{flushright}
}

\title{\Large Anomalies\hspace{-0.6pt} From\hspace{-0.6pt} the\hspace{-0.6pt} Covariant\hspace{-0.6pt} Derivative\hspace{-0.6pt} Expansion}

\author[1,2,3]{Timothy~Cohen,}
\author[4,3]{Xiaochuan~Lu,}
\author[5]{and Zhengkang~Zhang}

\affiliation[1]{\fontsize{9}{9}\selectfont \,Theoretical Physics Department, CERN, 1211 Geneva, Switzerland}
\affiliation[2]{\fontsize{9}{9}\selectfont \,Theoretical Particle Physics Laboratory, EPFL, 1015 Lausanne, Switzerland}
\affiliation[3]{\fontsize{9}{9}\selectfont \,Institute for Fundamental Science, University of Oregon, Eugene, OR 97403, USA}
\affiliation[4]{\fontsize{9}{9}\selectfont \,Department of Physics, University of California, San Diego, La Jolla, CA 92093, USA}
\affiliation[5]{\fontsize{9}{9}\selectfont \,Department of Physics, University of California, Santa Barbara, CA 91106, USA}

\emailAdd{tim.cohen@cern.ch}
\emailAdd{xil224@ucsd.edu}
\emailAdd{zkzhang@ucsb.edu}

\abstract{We revisit the calculation of anomalies for global and gauge symmetries in the framework of the Covariant Derivative Expansion (CDE). Due to the presence of UV divergences, the result is an ambiguous quantity that depends on the regularization procedure and the renormalization scheme. We introduce a class of regulators that facilitate a straightforward evaluation of the anomaly exclusively in $d=4$ spacetime dimensions using the CDE methodology. We derive a master formula for the anomaly that integrates various known results into a unified framework.}

\begin{document}
\maketitle
\flushbottom
\setcounter{page}{2}
\newpage

\begin{spacing}{1.1}
\parskip=0ex

%%%%%%%%%%%%%%%%%%%%%%%%%%%%%%%%%%%%%%%%%%%%%%%%%%%%%%%%%%%%%%%%%%%%%%%%%%%%%%%%
\section{Introduction}
\label{sec:Introduction}
%%%%%%%%%%%%%%%%%%%%%%%%%%%%%%%%%%%%%%%%%%%%%%%%%%%%%%%%%%%%%%%%%%%%%%%%%%%%%%%%

It is well known that symmetries of the classical action can be broken by quantum effects. This so-called anomaly has far-reaching consequences, from explaining the neutral pion decay to providing critical consistency checks on gauge theories with chiral fermions. Well-established techniques exist for computing anomalies both using Feynman diagrams, and also directly from the path integral. They can be computed for global and gauged symmetries, Abelian and non-Abelian groups, and take `consistent' and/or `covariant' forms. The results generally depend on the choice of regulator, and consist of a relevant piece that reflects the IR properties of the theory, and an irrelevant piece that can be absorbed by varying the renormalization scheme.\footnote{There is of course a vast literature on the anomaly, including the excellent reviews Refs.~\cite{Bertlmann:1996xk,Bilal:2008qx}. The story began with its discovery in 1969 by Adler~\cite{Adler:1969gk} and by Bell and Jackiw~\cite{Bell:1969ts}. It was soon after understood to be one-loop exact~\cite{Adler:1969er}. The connection between the anomaly and the topological winding number of the gauge field was discovered in Refs.~\cite{Jackiw:1976pf, Nielsen:1976hs, Nielsen:1977aw, Nielsen:1977qm}.  Of great importance to the approach taken here is Fujikawa's derivation of the anomaly from the non-invariance of the path integral measure~\cite{Fujikawa:1979ay, Fujikawa:1980eg, Fujikawa:1983bg, Fujikawa:1984pt, Fujikawa:2004cx}.}

In this paper, we revisit the calculation of anomalies from the path integral using an approach known as the Covariant Derivative Expansion (CDE).  This allows us to derive a unified framework that incorporates various types of anomalies into one master formula. The CDE was originally invented in the mid-1980s \cite{Gaillard:1985uh, Chan:1986jq, Cheyette:1987qz} to facilitate one-loop calculations of correlation functions purely in terms of functional traces, avoiding the introduction of Feynman diagrams. In recent years, the method has been applied in a variety of new settings, which has led to significant theoretical developments. These include the discovery of a variation on the framework, `simplified CDE'~\cite{Henning:2016lyp, Cohen:2019btp}, the incorporation of the method of regions~\cite{Fuentes-Martin:2016uol}, organizing schemes using diagrammatic frameworks \cite{Zhang:2016pja,Cohen:2020fcu}, as well as techniques that yield effective actions that include all orders in the fields \cite{Cohen:2020xca}. With these developments, the power and efficiency of CDE has been demonstrated for connecting the UV with the IR, \ie, computing low-energy Effective Field Theories (EFTs) from integrating out heavy states in a perturbative UV model; see \eg\ Ref.~\cite{Cohen:2022tir} for a review. We now know how to use CDE to perform matching calculations across a mass threshold, as well as to extract the renormalization group evolution equations for the EFT couplings. The CDE has become such a well-developed tool that there now exist packages which automate these calculations \cite{Cohen:2020qvb, Fuentes-Martin:2020udw, Fuentes-Martin:2022jrf}.

The practical success of CDE in connecting UV and IR descriptions of quantum field theories motivates applying it to compute the anomaly. The approach taken here will be to work exclusively in $d=4$ spacetime dimensions, which allows us to avoid any of the complications that arise when attempting to define Weyl fermions in dimensional regularization.\footnote{It is well known that handling the $\gamma^5$ matrix in $d\neq4$ spacetime dimensions is a nontrivial task \cite{tHooft:1972tcz, Breitenlohner:1977hr, Chanowitz:1979zu, Jegerlehner:2000dz}. See Refs.~\cite{Quevillon:2021sfz, Filoche:2022dxl} for recent CDE calculations of anomalies with dimensional regularization.}
In this paper, we generalize the classic Fujikawa approach by expressing the anomaly as a functional trace, which must be regularized to be well-defined. We introduce a novel regularization prescription, with a set of regulators parameterized by a set of numbers collectively denoted by $\beta$, which one can choose based on which symmetries one wishes to preserve. We emphasize that our regularization yields \emph{unambiguous} evaluation results once the values of $\beta$ are specified. We derive a master formula for the anomaly using CDE, whose explicit forms are given by \cref{eqn:Master1,eqn:Master2}. This master formula encodes a variety of known results for anomaly calculations. In particular, we examine all possible combinations of continuous symmetry groups, and show in each case how our master formula reproduces the known (relevant) anomaly results, as well as the anomaly cancellation conditions. This establishes that the CDE can accommodate this important effect in perturbative quantum field theory, and sets the stage for its applications to EFT matching across anomalous thresholds.

The rest of this paper is organized as follows. We first review the functional formalism in \cref{sec:PathIntegral}, with an emphasis on the definition of the anomaly and its connections to the fermionic path integral measure and the anomalous Ward identities. In \cref{sec:Regularization}, we isolate the functional trace that encodes the anomalies and introduce our novel regularization prescription to make it well-defined. We discuss the relation between our regulator and some similar approaches in the literature, and also a sufficient condition for it to be consistent with the Wess-Zumino condition. In \cref{sec:Evaluation}, we carry out the CDE evaluation to obtain our master formula for the anomaly. We then demonstrate in \cref{sec:Application} that this master formula reproduces various known results regarding anomalies by examining all possible combinations of continuous symmetry groups. Some future directions are discussed in \cref{sec:Discussion}. A technical clarification regarding CDE manipulations is provided in \cref{appsec:perm}.

%%%%%%%%%%%%%%%%%%%%%%%%%%%%%%%%%%%%%%%%%%%%%%%%%%%%%%%%%%%%%%%%%%%%%%%%%%%%%%%%
\section{Anomalies in the Functional Formalism}
\label{sec:PathIntegral}
%%%%%%%%%%%%%%%%%%%%%%%%%%%%%%%%%%%%%%%%%%%%%%%%%%%%%%%%%%%%%%%%%%%%%%%%%%%%%%%%

In this section, we briefly review the well-known functional formalism for anomalies, which also serves the purpose of introducing our notation. Much of this section is drawn from the review article by Bilal~\cite{Bilal:2008qx}. Our discussion here crucially relies on the famous connection between anomalies and the path integral measure first discovered by Fujikawa~\cite{Fujikawa:1979ay}.

\subsection{Defining the Anomaly}
\label{subsec:Def} 

We begin with the definition of the anomaly. Consider a general gauge theory coupled to a set of left-handed Weyl fermions collectively denoted by $\chi$: 
\begin{equation}
\L = -\frac{1}{4g^2} F_{\mu\nu}^a F^{a\mu\nu} + \chi^\dagger \bar\sigma^\mu P_\mu \chi \,,
\label{eqn:LagMinimal}
\end{equation}
where we have defined the Hermitian covariant derivative
\begin{equation}
P_\mu \equiv iD_\mu = i\partial_\mu + G_\mu = i\partial_\mu +G_\mu^a t^a \,,
\label{eqn:PmuDef}
\end{equation}
where $t^a$ are the (Hermitian) gauge group generators. The gauge field strength is given by
\begin{equation}
F_{\mu\nu} = F_{\mu\nu}^a\, t^a = -i\, \comm{P_\mu}{P_\nu} = \left(\partial_\mu G_\nu\right) - \left(\partial_\nu G_\mu\right) - i\, \comm{G_\mu}{G_\nu} \,.
\label{eqn:FmunuDef}
\end{equation}
The kinetic term for the gauge fields in \cref{eqn:LagMinimal} should be read as a sum over terms normalized with different gauge couplings in the case of a product gauge group.

A gauge transformation can be parameterized by the matrix
\begin{equation}
U_\alpha = e^{i\alpha} = e^{i\alpha^a t^a} \,,
\label{eqn:UDef}
\end{equation}
where the transformation parameters $\alpha^a = \alpha^a(x)$ are functions of spacetime. Under \cref{eqn:UDef}, the building blocks of our theory transform as
\begin{subequations}\label{eqn:GaugeTransformation}
\begin{align}
\chi &\quad\to\quad \chi_\alpha = 
U_\alpha \chi \,, \\[8pt]
\chi^\dagger &\quad\to\quad \chi_\alpha^\dagger = 
\chi^\dagger U_\alpha^\dagger \,, \\[8pt]
P^\mu &\quad\to\quad P^\mu_\alpha = 
U_\alpha P^\mu U_\alpha^\dagger \,, \\[8pt]
G^\mu &\quad \to \quad G^\mu_\alpha = 
U_\alpha G^\mu U_\alpha^\dag +  U_\alpha \left( i\partial^\mu U_\alpha^\dagger \right) \,. \label{eqn:GTransformation}
\end{align}
\end{subequations}
We will use $\delta_\alpha$ to denote the first-order (in $\alpha$) gauge variation; for example,
\begin{equation}
\delta_\alpha G^\mu \equiv (G^\mu_\alpha - G^\mu)\bigr|_{\O(\alpha)} = D^\mu \alpha = \partial^\mu \alpha -i\bigl[G^\mu, \alpha \bigr] \,.
\end{equation}

The Lagrangian in \cref{eqn:LagMinimal} defines an action that is gauge invariant at the classical level. However, quantum effects can spoil gauge invariance. If this happens, we say that the theory has an anomaly.

To define the anomaly, we consider the bosonic effective action $W[G]$, computed from the path integral by integrating out the fermions, while treating the gauge field as a classical background:
\begin{equation}
e^{iW[G]} \equiv \int \Dcal\chi\s \Dcal\chi^\dagger\, e^{iS_\text{f}[\chi,\chi^\dagger, G]} \,,
\label{eqn:EffectiveAction}
\end{equation}
where $S_\text{f}\equiv \int \dd^4 x\, \chi^\dagger \bar\sigma^\mu P_\mu \chi$ is the fermion bilinear part of the classical action. If we would also like to treat the gauge field $G_\mu$ as a dynamical quantum field by performing its path integral,
\begin{equation}
\int \Dcal G\s \Dcal\chi\s \Dcal\chi^\dagger\, e^{i\int \dd^4x \,\L} = \int \Dcal G \, e^{\,i\bigl(-\frac{1}{4g^2}\int \dd^4x \,F^a_{\mu\nu} F^{a\mu\nu} + W[G]\bigr)} \,,
\label{eqn:PathIntegralG}
\end{equation}
we need $W[G]$ to be gauge invariant (upon regularization and renormalization). Gauge invariance of the classical action $S_\text{f}$ does not guarantee that of $W[G]$, since quantum effects (due to the fermionic path integral measure) can break gauge invariance.

The anomaly functional $\Acal [\alpha]$, which we also simply refer to as the anomaly, can be defined by taking the gauge variation of the bosonic effective action $W[G]$:\footnote{Note that in such variations, we restrict to the set of $\alpha(x)$ that fall off fast enough at infinity such that one can always use integration by parts (see \eg\ \cref{eqn:WGvariation} below). In particular, a constant $\alpha(x)$ does not belong to this set.}
\begin{equation}
\Acal [\alpha] \equiv \int \dd^4x\, \alpha^a(x) \Acal^a(x) \equiv \delta_\alpha W[G] \,.
\label{eqn:AnomalyDef}
\end{equation}
If $\Acal [\alpha]=0$, the theory is anomaly-free and the path integral in \cref{eqn:PathIntegralG} yields a well-behaved quantum theory. If $\Acal [\alpha]\ne0$ but is equal to the gauge variation of a local action, $\Acal [\alpha] = \delta_\alpha (-\int \dd^4x \,\L_\text{ct})$, it is called an {\it irrelevant} anomaly and can be removed by renormalization, \ie, by adding local counterterms $\L_\text{ct}$ to the Lagrangian (see \eg\ Ref.~\cite{Cornella:2022hkc} for a systematic study of such counterterms); in this case the (renormalized) quantum theory is also well-behaved. On the other hand, a nonzero $\Acal[\alpha]$ that cannot be written as the gauge variation of a local action, called a {\it relevant} anomaly, implies that the gauge theory is not well-defined at the quantum level; in this case, the anomaly is an IR effect and cannot be removed by renormalization.

The definition \cref{eqn:AnomalyDef} we adopt here is known as the \emph{consistent} anomaly, in the sense that it should -- if properly regularized -- satisfy the Wess-Zumino consistency condition \cite{Wess:1971yu}:
\begin{equation}
\delta_{\alpha_1} \Acal[\alpha_2] - \delta_{\alpha_2} \Acal[\alpha_1] =  \Acal\bigl[-i[\alpha_1,\alpha_2]\bigr] \,,
\end{equation}
which is a direct consequence of the Lie algebra:
\begin{equation}
( \delta_{\alpha_1} \delta_{\alpha_2} - \delta_{\alpha_2} \delta_{\alpha_1} ) W[G] = \delta_{-i[\alpha_1,\alpha_2]} W[G] \,. 
\end{equation}
The Wess-Zumino consistency condition is also equivalent to the statement that the anomaly is BRST-closed when $\alpha$ is replaced by the ghost field $\omega=\omega^at^a$:
\begin{equation}
\Acal[\omega] = \delta_\brst W[G]
\qquad\Longrightarrow\qquad
\delta_\brst \Acal[\omega] = 0 \,,
\label{eqn:AcalBRST}
\end{equation}
which follows from the nil-potency of the BRST transformation, $\delta_\brst^2=0$. However, since the bosonic effective action $W[G]$ is not a local functional of the gauge field $G_\mu$, the fact that $\Acal[\omega] = \delta_\brst W[G]$ does not mean that the anomaly is BRST-exact on the space of local functionals. Anomalies that are BRST-exact on this space can be absorbed by local counterterms, $\Acal[\omega] = \delta_\brst (-\int \dd^4x \,\L_\text{ct})$, and are the irrelevant anomalies, while the relevant anomalies are given by nontrivial BRST cohomology classes (closed but not exact) on this space.

Finally, we note that while we have focused on gauge symmetries in the discussion above, anomalies of global symmetries can be treated in the same framework by artificially gauging all the (classical) global symmetries of interest. Concretely, we introduce auxiliary gauge fields for all the global symmetries as part of $G_\mu$, and take $U_\alpha$ to also include local transformations associated with the global symmetry generators. Then $\Acal[\alpha]$ as defined above will also contain anomalies of the global symmetries, and a nonzero value of $\Acal[\alpha]$ implies that the classical global symmetry cannot be gauged in the quantum theory. In what follows, we will assume this artificial gauging has been done for all the classical global symmetries of interest, and will not distinguish between global and gauge symmetries.

\subsection{Connection to the Path Integral Measure}
\label{subsec:Measure}

As explained above, the classical action $S_\text{f}$ in \cref{eqn:EffectiveAction} is gauge invariant, so the only possible source of the anomaly is the path integral measure over the fermionic fields. Specifically, performing the transformation in \cref{eqn:GaugeTransformation} changes the measure by a Jacobian factor:
\begin{equation}
\Dcal\chi_\alpha \Dcal\chi_\alpha^\dagger = \Jcal_\alpha^{-1}\s \Dcal\chi\s \Dcal\chi^\dagger \,.
\label{eqn:Jacobian}
\end{equation}
Therefore, we have
\begin{align}
e^{iW[G_\alpha]} &= \int \Dcal\chi\s \Dcal\chi^\dagger\, e^{iS_\text{f}[\chi, \chi^\dagger, G_\alpha]}
= \int \Dcal\chi_\alpha \Dcal\chi_\alpha^\dagger\, e^{iS_\text{f}[\chi_\alpha, \chi_\alpha^\dagger, G_\alpha]} \notag\\[5pt]
&
= \int \Jcal_\alpha^{-1} \Dcal\chi\s \Dcal\chi^\dagger\, e^{iS_\text{f}[\chi, \chi^\dagger, G]}
\notag\\[5pt]
&= e^{iW[G]}\, \frac{\int \Jcal_\alpha^{-1} \Dcal\chi\s \Dcal\chi^\dagger\, e^{iS_\text{f}[\chi, \chi^\dagger, G]}}{\int \Dcal\chi\s\Dcal\chi^\dagger\, e^{iS_\text{f}[\chi, \chi^\dagger, G]}}
= e^{iW[G]} \left\langle \Jcal_\alpha^{-1}\right\rangle_G \,.
\label{eqn:WJacobian}
\end{align}
In the first line, we just relabeled the dummy integration variables, $\chi \to \chi_\alpha$; in the second line, we used \cref{eqn:Jacobian} and the gauge invariance of the classical action $S_\text{f}$; in the last line, we multiplied and divided the expression by $e^{iW[G]}=\int \Dcal\chi\s\Dcal\chi^\dagger\, e^{iS_\text{f}[\chi, \chi^\dagger, G]}$. Taking the logarithm of \cref{eqn:WJacobian}, we arrive at a relation between the Jacobian factor\footnote{We note that sometimes in the literature, $\left\langle \Jcal_\alpha^{-1}\right\rangle_G$ is simply written as $\Jcal_\alpha^{-1}(G)$ or just $\Jcal_\alpha^{-1}$. This might give an impression that it does not depend on the details of the action $S_\text{f}$. Throughout this paper, we manifestly write it as an expectation value $\left\langle \Jcal_\alpha^{-1}\right\rangle_G$ to emphasize that it is a quantum expectation value and \emph{a priori} may depend on what interactions are included in the action.} and the anomaly:
\begin{equation}
-i \log \left\langle \Jcal_\alpha^{-1}\right\rangle_G = W[G_\alpha] - W[G] = \Acal[\alpha] + \O(\alpha^2) \,.
\end{equation}
We see that when the quantum expectation value of the Jacobian factor is trivial, there is no anomaly
\begin{equation}
\left\langle \Jcal_\alpha^{-1}\right\rangle_G = 1
\qquad\Longrightarrow\qquad \Acal[\alpha]= 0 \,,
\label{eqn:AcalJcalG}
\end{equation}
while anomalies are associated with the quantum breaking of classical symmetries.

\subsection{Connection to Ward Identities}
\label{subsec:Ward}

The connection between the anomaly and the Ward identities can be made by noting
\begin{align}
\delta_\alpha W[G] &= \int \dd^4x\, \big[ \delta_\alpha G_\mu^a(x) \big] \frac{\delta W}{\delta G_\mu^a(x)}
= \int \dd^4x\,  (D_\mu \alpha)^a \frac{\delta W}{\delta G_\mu^a(x)} 
\notag\\[5pt]
&= -\int \dd^4x\, \alpha^a(x) \bigg[D_\mu \frac{\delta W}{\delta G_\mu(x)}\bigg]^a \,.
\label{eqn:WGvariation}
\end{align}
Comparing this to \cref{eqn:AnomalyDef}, we get
\begin{equation}
\left[D_\mu \frac{\delta W}{\delta G_\mu(x)}\right]^a = - \Acal^a(x) \,.
\label{eqn:deltaWAcal}
\end{equation}
Meanwhile, since $G_\mu^a$ acts as a source for the fermion current $J^{a\mu}=\chi^\dagger \bar\sigma^\mu t^a\chi$, we have
\begin{equation}
\frac{\delta W}{\delta G_\mu^a(x)} = \left\langle J^{a\mu} \right\rangle_G \,.
\end{equation}
Together, these imply
\begin{equation}
\left( D_\mu \left\langle J^\mu \right\rangle_G \right)^a = - \Acal^a(x) \,,
\label{eqn:CurrentAcal}
\end{equation}
\ie\ the fermion current is covariant up to the anomaly. The BRST symmetry that is critical to the quantization of gauge theory requires $\left( D_\mu \left\langle J^\mu \right\rangle_G \right)^a = 0$. This makes the connection between anomaly cancellation and consistency of gauge theory precise.

We can use \cref{eqn:CurrentAcal}, or equivalently \cref{eqn:deltaWAcal}, as a generating functional for the Ward identities. First, let us explicitly write out  the left-hand side of \cref{eqn:deltaWAcal}:
\begin{equation}
\partial_\mu \left( \frac{\delta W}{\delta G_\mu^a} \right) + f^{abc}\, G_\mu^b\, \frac{\delta W}{\delta G_\mu^c} = - \Acal^a(x) \,.
\end{equation}
Now taking the $k^\text{th}$ functional derivative, we get
\begin{align}
&\partial_\mu \left( 
\frac{\delta^{k+1} W}{\delta G_\mu^a \delta G_{\mu_1}^{b_1} \cdots \delta G_{\mu_k}^{b_k}} \right) \Biggr|_{G=0} + \sum_{i=1}^k f^{ab_ic} 
\frac{\delta^k W}{\delta G_{\mu_1}^{b_1} \cdots \delta G_{\mu_i}^c \cdots \delta G_{\mu_k}^{b_k}} \Biggr|_{G=0}
\notag\\[7pt]
&\hspace{100pt}
= - \frac{\delta^k \Acal^a(x)}{\delta G_{\mu_1}^{b_1} \cdots \delta G_{\mu_k}^{b_k}} \Biggr|_{G=0} \,.
\label{eqn:WardIdentities1}
\end{align}
These are the anomalous Ward identities, and are often written in terms of the connected correlation functions of the fermion currents:
\begin{align}
&\partial_\mu \langle J^{\mu, a} J^{\mu_1, b_1} \cdots J^{\mu_k, b_k} \rangle_\text{conn}
+ \sum_{i=1}^k f^{ab_ic} \langle J^{\mu_1, b_1} \cdots J^{\mu_i, c} \cdots J^{\mu_k b_k} \rangle_\text{conn}
\notag\\[5pt]
&\hspace{100pt}
= - \frac{\delta^k \Acal^a(x)}{\delta G_{\mu_1}^{b_1} \cdots \delta G_{\mu_k}^{b_k}} \Biggr|_{G=0} \,.
\label{eqn:WardIdentities2}
\end{align}
We see that a $G^k$ term in $\Acal^a(x)$ corresponds to a mismatch between the $(k+1)$-point and $k$-point correlation functions of the fermion currents. \cref{eqn:WardIdentities2} is sometimes taken as a definition of the anomaly in renormalized perturbation theory. In the case of an irrelevant anomaly, one can add local counterterms which give additional contributions to the left-hand side of \cref{eqn:WardIdentities1} and correspond to choosing a different renormalization scheme for the current correlators in \cref{eqn:WardIdentities2}. A relevant anomaly, on the other hand, constitutes a genuine violation of the classical Ward identities that cannot be remedied by renormalization. It is also worth noting that \cref{eqn:WardIdentities2} can be used to prove that $\Acal^a(x)$ truncates at a finite power of the gauge field $G_\mu$.

%%%%%%%%%%%%%%%%%%%%%%%%%%%%%%%%%%%%%%%%%%%%%%%%%%%%%%%%%%%%%%%%%%%%%%%%%%%%%%%%
\section{Regularizing the Anomaly}
\label{sec:Regularization}
%%%%%%%%%%%%%%%%%%%%%%%%%%%%%%%%%%%%%%%%%%%%%%%%%%%%%%%%%%%%%%%%%%%%%%%%%%%%%%%%

The definition in \cref{eqn:AnomalyDef} does not fully specify the value of the anomaly, because (the gauge variation of) the bosonic effective action $W[G]$ is not well-defined in the absence of a regulator. In this section, we introduce our regularization prescription. Then the CDE evaluation of the regularized anomaly will be presented in \cref{sec:Evaluation}.

Before discussing the case of anomalies, we first review the basic idea of regularization and illustrate the role of regularization prescriptions in \cref{subsec:Comments} using some simple toy series. (Experts can safely skip this subsection.) Then in \cref{subsec:beta}, we introduce our regularization prescription for the anomaly, motivated by its convenience for evaluating the functional trace using CDE. Specifically, we will be working in strictly $d=4$ spacetime dimensions, \ie, we will not be using dimensional regularization. Instead, we will insert a damping factor into the functional trace, in a similar spirit to heat kernel regularization. In fact, we will introduce a class of such damping factors parameterized by a set of numbers $\beta$; different choices of these $\beta$ parameters correspond to different regularization schemes and will lead to different results. In \cref{subsec:Connections}, we comment on the connection between our regularization prescription and some familiar approaches in the literature. In particular, we will see that both the heat kernel and Pauli-Villars regulators can be viewed as specific incarnations of our approach. Finally, we check our regularization prescription against the Wess-Zumino consistency condition in \cref{subsec:WessZumino}, and show how it may be satisfied or violated depending on the choice of $\beta$ values.

\subsection{What is Regularization?}
\label{subsec:Comments}

In this subsection, we illustrate the role of regularization with some simple toy series. In particular, we demonstrate how different regularization prescriptions correspond to different definitions for a non-converging series and hence generically lead to different results upon evaluation. We will also clarify the allowed manipulations for a non-converging series.

When we encounter a series that is not convergent, its sum does not have a well-defined value. However, it is often useful to promote such a series into a `function series,' where the summands are functions; these functions must reproduce the original series term by term when their argument takes a particular value (or limit). Then we can define the sum through analytic continuation: we first sum the function series inside its convergence region to obtain an analytic function, and then take the limit corresponding to the original series to define the value of the latter. This regularization procedure leads to a regulated (finite) series.

Let us explain how this works using a simple example. Consider the series
\begin{equation}
s_1 = \sum_{k=0}^\infty 2^k = 1 + 2 + 4 + 8 + \cdots \,.
\end{equation}
Clearly, this is a non-converging series. However, we could associate it with the function series
\begin{equation}
s_1 \quad\Longleftrightarrow\quad
\left( \sum_{k=0}^\infty x^k \right) \Biggr|_{x=2}
\quad\xrightarrow{\text{ regularization }}\quad
\frac{1}{1-x} \biggr|_{x=2} = -1 \,.
\end{equation}
This function series converges to $f_1(x) = \frac{1}{1-x}$ within the disk $|x|<1$, but not at $x=2$. But we can take $f_1(x=2)$ as the definition for the sum $s_1$. This is what we mean by a regulated series. Another example is the famous zeta function regularization originally used by Euler: the diverging series
\begin{equation}
s_2 = \sum_{k=1}^\infty k = 1 + 2 + 3 + 4 + \cdots 
\end{equation}
can be regularized as
\begin{equation}
s_2 \quad\Longleftrightarrow\quad
\left( \sum_{k=1}^\infty \frac{1}{k^s}\right)\Biggr|_{s=-1}
\quad\xrightarrow{\text{ regularization }}\quad
\zeta(s)\bigr|_{s=-1} = -\frac{1}{12} \,.
\end{equation}

As mentioned above, when we promote a non-converging series into a function series, we require that the function series  reproduces the original series term by term when evaluated at a certain point. Clearly, this does not uniquely specify the choice: given a non-converging series, one can usually promote it into many different function series. These correspond to different regularization schemes and serve as different definitions of the sum of the original series. To see this concretely, let us consider the following toy series
\begin{equation}
s_0 = \sum_{k=0}^\infty (-1)^k = 1 - 1 + 1 - 1 + 1 - 1 + \cdots \,.
\end{equation}
To regularize this series, we could choose to promote it to any of the following set of function series parameterized by a number $\beta$:
\begin{equation}
f_\beta (\tau) = \tau^0 - \tau^{1+\beta} + \tau^2 - \tau^{3+\beta} + \tau^4 - \tau^{5+\beta} + \cdots \,.
\label{eqn:fbetatau}
\end{equation}
Then we have
\begin{equation}
s_0 \quad\Longleftrightarrow\quad
f_\beta (\tau) \bigr|_{\tau\to1}
\quad\xrightarrow{\text{ regularization }}\quad
\frac{1-\tau^{1+\beta}}{1-\tau^2} \Biggr|_{\tau\to1} = \frac{1+\beta}{2} \,.
\label{eqn:fbetareg}
\end{equation}
We see that with different values for $\beta$, the original non-converging series $s_0$ can be defined/regularized to take different values.

If we are going to regularize a non-converging series with a function series that is absolutely convergent (in its convergence region), then one can shuffle and/or group terms in the latter without changing its analytic continuation. Alternatively, one could shuffle and/or group terms first in the original non-converging series, and then regularize the new expression with an absolutely convergent function series. This second way will lead to the same  result upon evaluation, and it is sometimes more convenient because the series is easier to massage before promoting it into a function series. However, when we shuffle and/or group terms in the original non-converging series to go from one expression to another, we have to remember that none of these expressions is well-defined yet, so it is not appropriate to say that they are equal (`$=$'). Instead, they are just `equivalent' in the sense that they would be equal if one were to regularize them with the \emph{same} absolutely convergent function series (with the same shuffling and/or grouping of terms). In this paper, we use the symbol `$\seq$' to denote this equivalence relation between non-converging series (see equations below starting from \cref{eqn:WGExpand}).

Let us take the same toy series example $s_0$ to illustrate this point, as well as the use of the `$\seq$' notation. Since the function series \cref{eqn:fbetatau} is absolutely convergent within the disk $|\tau|<1$, we can group its terms to get another series:
\begin{equation}
f_\beta (\tau) \xrightarrow{\text{ group terms }} {\widetilde f}_\beta (\tau) \equiv \sum_{k=0}^\infty \left( \tau^{2k} - \tau^{2k+1+\beta} \right) \xrightarrow{\text{ analytic continuation }} \frac{1-\tau^{1+\beta}}{1-\tau^2} \,,
\end{equation}
which has the same analytic continuation. Alternatively, one could first group terms in the original non-converging series:
\begin{equation}
s_0 \,\seq\, {\widetilde s}_0 \equiv \sum_{k=0}^\infty (1-1) = 0 + 0 + 0 + \cdots \,.
\end{equation}
Note that we have used the `$\seq$' sign here between $s_0$ and $\widetilde s_0$. The new series $\widetilde s_0$ is a converging series and does have a default definition, so a regularization for $\widetilde s_0$ is not mandatory. However, one could still use the function series $\widetilde f_\beta (\tau)$ to regularize it, because
\begin{equation}
\left( \tau^{2k} - \tau^{2k+1+\beta} \right) \Bigr|_{\tau = 1} = 0 
\end{equation}
would also reproduce the series $\widetilde s_0$ term by term. With this regularization, one would then get the same evaluation result $\frac{1+\beta}{2}$ as in \cref{eqn:fbetareg}. Our use of the `$\seq$' sign here is emphasizing this: $s_0$ and $\widetilde s_0$ are equal only when we use the same regularization prescription for them (although one of them has a different default definition in the absence of regularization).

We note in particular that performing cyclic permutations inside a trace is a typical type of shuffling and/or grouping of terms:
\begin{subequations}
\begin{align}
\tr\, (AB) &= \sum_i \left( \sum_a A_{ia} B_{ai} \right) \,, \\[5pt]
\tr\, (BA) &= \sum_a \left( \sum_i B_{ai} A_{ia} \right) = \sum_a \left( \sum_i A_{ia} B_{ai} \right) \,.
\end{align}
\end{subequations}
The two traces are related by a change of summation order. When the matrices $A$ and $B$ are infinite dimensional, such as in the case of functional traces, each trace is a sum over a (double) series. If the series is not convergent and needs regularization to be well-defined, then it is not appropriate to claim that the two traces are equal, as we have just explained. Instead, we should use the `$\seq$' sign:
\begin{equation}
\tr\, (AB) \seq \tr\, (BA) \,,
\end{equation}
to emphasize that they would be equal when we use the same absolutely convergent function series to regulate them.

\subsection{Anomaly as a Regulated Functional Trace}
\label{subsec:beta}

Let us now turn to the case of interest in this paper, the anomaly functional $\Acal[\alpha]$ defined in \cref{eqn:AnomalyDef}. First, we would like to isolate the functional trace that encodes the anomalies. We start with the definition of $W[G_\alpha]$, \cref{eqn:EffectiveAction} with $G^\mu$ replaced by $G^\mu_\alpha$ according to \cref{eqn:GTransformation}. It can be formally written as a functional determinant:
\begin{equation}
e^{iW[G_\alpha]} = \int \Dcal\chi\s \Dcal\chi^\dagger\, e^{iS_\text{f}[\chi, \chi^\dagger, G_\alpha]}
= \det \left( U_\alpha\, \bar\sigma^\mu P_\mu\, U_\alpha^\dagger \right) \,.
\end{equation}
Taking the logarithm and expanding in $\alpha$, we get
\begin{align}
W[G_\alpha] &= -i\log\det \left( U_\alpha\, \bar\sigma^\mu P_\mu\, U_\alpha^\dagger \right)
\notag\\[10pt]
&\seq -i\log\det \left( \bar\sigma^\mu P_\mu + i\alpha\, \bar\sigma^\mu P_\mu - \bar\sigma^\mu P_\mu\, i\alpha \right) + \O(\alpha^2) 
\notag\\[5pt]
&\seq -i\log\det \left( \bar\sigma^\mu P_\mu \right) -i \log\det \left[ 1 + \frac{1}{\bar\sigma^\nu P_\nu}\, \left( i\alpha\, \bar\sigma^\mu P_\mu - \bar\sigma^\mu P_\mu\, i\alpha \right) \right] + \O(\alpha^2)
\notag\\[5pt]
&\seq W[G]-i \Tr\log \left[ 1 + \frac{1}{\bar\sigma^\nu P_\nu}\, \left( i\alpha\, \bar\sigma^\mu P_\mu - \bar\sigma^\mu P_\mu\, i\alpha \right) \right] + \O(\alpha^2)
\notag\\[5pt]
&\seq W[G] + \Tr \left[ \frac{1}{\bar\sigma^\nu P_\nu}\, \left( \alpha\, \bar\sigma^\mu P_\mu - \bar\sigma^\mu P_\mu\, \alpha \right) \right] + \O(\alpha^2) \,.
\label{eqn:WGExpand}
\end{align}
According to the definition in \cref{eqn:AnomalyDef}, the leading order contribution to the difference $W[G_\alpha]-W[G]$ gives the anomaly. Therefore, we obtain
\begin{equation}
\Acal[\alpha] \seq \Tr \left[ \frac{1}{\bar\sigma^\nu P_\nu}\, \bigl( \alpha\, \bar\sigma^\mu P_\mu - \bar\sigma^\mu P_\mu\, \alpha \bigr) \right] \,.
\label{eqn:AnomalyTraceSigma}
\end{equation}

At this point, we have formally written the anomaly as a functional trace. However, we emphasize that the functional trace in \cref{eqn:AnomalyTraceSigma} is the sum of a series that is not convergent, so it does not have a definite value and requires regularization to become well-defined. As elaborated in \cref{subsec:Comments}, different regularization prescriptions can yield different results upon evaluation. In fact, the same is true for the expression in each line of \cref{eqn:WGExpand}. Therefore, we have used the notation `$\seq$' to emphasize that these expressions are not exactly equal `$=$' unless they are regularized in the same way.

One may also attempt to perform a cyclic permutation within the functional trace in \cref{eqn:AnomalyTraceSigma}, so the two terms appear to cancel. However, as explained in \cref{subsec:Comments}, such a cyclic permutation amounts to shuffling and/or grouping terms in the original non-converging series to obtain a new series:
\begin{equation}
\Acal[\alpha] \seq \Tr[0] = 0 + 0 + 0 + \cdots \,.
\label{eqn:zerossum}
\end{equation}
Although this new series is zero term by term, which is convergent and hence has a default definition without a regulator, this does not contradict the statement that regularization prescriptions exist that yield a nonzero value for this series. As emphasized by the `$\seq$' sign, the two expressions above would only be equal under the same regularization prescription. The default definition of the right-hand side (which gives zero) corresponds to one particular choice of regularization (a trivial one), so its evaluation result would not be equal to that of the left-hand side if a different regularization prescription is chosen for the latter.

To motivate our regulator, let us first check what would happen if we go ahead and evaluate the functional trace in \cref{eqn:AnomalyTraceSigma} with CDE. Focusing on the first term, we have
\begin{align}
\Tr \left( \frac{1}{\bar\sigma^\nu P_\nu}\, \alpha\, \bar\sigma^\mu P_\mu \right) &\seq
\int \dd^4 x \int \frac{\dd^4 q}{(2\pi)^4} \tr\left[ \frac{1}{\bar\sigma^\nu (q_\nu + P_\nu)}\, \alpha\, \bar\sigma^\mu (q_\mu + P_\mu) \right]
\notag\\[8pt]
&\hspace{-40pt}
\seq \int \dd^4 x \int \frac{\dd^4 q}{(2\pi)^4} \tr\left[ \sum_{k=0}^\infty \left( -\frac{\sigma^\lambda q_\lambda}{q^2}\, \bar\sigma^\tau P_\tau \right)^k \frac{\sigma^\nu q_\nu}{q^2}\, \alpha\, \bar\sigma^\mu (q_\mu + P_\mu) \right]
\notag\\[8pt]
&\hspace{-40pt}
\seq \int \dd^4 x \int \frac{\dd^4 q}{(2\pi)^4} \tr\left[ \sum_{k=0}^\infty \left( -\frac{\qsl}{q^2}\, \Psl \right)^k \frac{\qsl}{q^2}\, \alpha\, (\qsl + \Psl)\, \frac{1-\gamma^5}{2} \right]
\notag\\[8pt]
&\hspace{-40pt}
\seq \int \dd^4 x \int \frac{\dd^4 q}{(2\pi)^4} \tr\left[ \sum_{k=0}^\infty \left( -\frac{\qsl}{q^2}\, \Pb{} \right)^k \frac{\qsl}{q^2}\, \alpha\, (\qsl + \Pb{})\, \frac{1-\gamma^5}{2} \right]
\notag\\[8pt]
&\hspace{-40pt}
\seq \int \dd^4 x \int \frac{\dd^4 q}{(2\pi)^4} \tr\left[ \frac{1}{\qsl + \Pb{}}\, \alpha\, (\qsl + \Pb{})\, \frac{1-\gamma^5}{2} \right]
\notag\\[8pt]
&\hspace{-40pt}
\seq \Tr \left( \frac{1}{\Pb{}}\, \alpha\, \Pb{}\, \frac{1-\gamma^5}{2} \right) \,.
\label{eqn:betaIntro}
\end{align}
The first line above simply follows from the definition of the functional trace.\footnote{See \cref{eqn:TrfCDE} for a more detailed explanation of the shift $P_\mu \to q_\mu + P_\mu$. We note that the internal traces `$\tr$' from here on in the main text are actually what we denote by `$\ttr$' in \cref{appsec:perm}. See \cref{appsubsec:trNotations}, especially the discussion around \cref{eqn:ttrDifferential} for a careful clarification on this notation.}
To obtain the second line, we have performed a Taylor expansion in terms of the Hermitian covariant derivative $P_\mu$, an operation called the Covariant Derivative Expansion (CDE) in the literature.\footnote{More precisely, the operation here is called `simplified CDE' \cite{Henning:2016lyp}, in which one makes a Taylor expansion directly in terms of the `open' covariant derivatives. This is different from the `original CDE' \cite{Gaillard:1985uh, Chan:1986jq, Cheyette:1987qz} where one inserts additional factors to `close' the covariant derivatives (\ie\ put them into commutators) before performing the Taylor expansion. See the discussion around \cref{eqn:openvsclosed} for an elaboration on open vs.\ closed derivatives in functional operators, and App.~B of Ref.~\cite{Cohen:2019btp} for a detailed discussion on simplified vs.\ original CDE.}
To get the third line, we used the following identity between Pauli matrices and the Dirac gamma matrices:
\begin{equation}
\tr \Big[ \left( \sigma^{\mu_1} \bar\sigma^{\nu_1} \right) \cdots \left( \sigma^{\mu_k} \bar\sigma^{\nu_k} \right) \Big] = \tr \left[ \left( \gamma^{\mu_1} \gamma^{\nu_1} \right) \cdots \left( \gamma^{\mu_k} \gamma^{\nu_k} \right) \frac{1-\gamma^5}{2} \right] \,.
\label{eqn:SigmaToGamma}
\end{equation}
Starting from the fourth line of \eqref{eqn:betaIntro}, we have introduced the $\beta$-modified covariant derivative:\footnote{Note that when $\beta\ne1$, the operator $\Pb{}$ is not gauge covariant. This is the reason why we will not always get a covariant anomaly; see discussion in \cref{subsec:WessZumino} for more details.}
\begin{align}
\Pb{} &\equiv i\pdsl + \Gsl \left( \frac{1-\gamma^5}{2} + \beta \frac{1+\gamma^5}{2} \right) 
\notag\\[8pt]
&\equiv i\pdsl + \sum_a \Gsl^a t^a \left( \frac{1-\gamma^5}{2} + \beta_a \frac{1+\gamma^5}{2} \right) \,.
\label{eqn:PslbetaDef}
\end{align}
Finally, in the last line of \cref{eqn:betaIntro}, we rewrote the result as a functional trace. Note that we take the $\beta_a$ parameters to be degenerate within each simple gauge group sector so that $\beta_a t^a$ (no sum over $a$) satisfy the same Lie algebra as $t^a$. Here and in what follows, we explicitly write out the summation over adjoint indices when the presence of $\beta_a$ results in more than two identical adjoint indices in an expression.

The identity in \cref{eqn:SigmaToGamma} has allowed us to convert the two-component expression (left-hand side of \cref{eqn:betaIntro}) into a four-component expression (last line in \cref{eqn:betaIntro}) with an insertion of the projector operator $\frac{1-\gamma^5}{2}$. The same procedure goes through when both terms in \cref{eqn:AnomalyTraceSigma} are present, so we have
\begin{equation}
\Acal[\alpha] \seq \Tr \left[ \frac{1}{\Pb{}} \left( \alpha\, \Pb{} - \Pb{}\, \alpha \right) \frac{1-\gamma^5}{2} \right] \,.
\label{eqn:Tbeta}
\end{equation}

At this stage, it seems that the $\beta$ parameters could take arbitrary values without affecting the value of the expression, because it comes with the factor $\frac{1+\gamma^5}{2}$ which will get annihilated by the projector $\frac{1-\gamma^5}{2}$ at the end of the expression. However, we stress that this $\beta$-parameterized functional trace is still the sum of a non-converging series, so we need to introduce a regulator to make it well-defined. As we will see below, once we regulate this expression, different $\beta$'s will \emph{define} different values for the functional trace.

Motivated by the form of the expression in \cref{eqn:Tbeta}, we choose to insert a damping factor to define the \emph{regularized} anomaly:
\begin{equation}
\Acal_\beta^\Lambda [\alpha] \equiv \Tr \left[ f\left(-\frac{\Pb{2}}{\Lambda^2}\right) \frac{1}{\Pb{}}\, \left( \alpha\, \Pb{} - \Pb{}\, \alpha \right) \frac{1-\gamma^5}{2} \right] \,,
\label{eqn:AcalRegularized}
\end{equation}
where the function $f(u)$ satisfies the following conditions:
\begin{subequations}\label{eqn:fconditions}
\begin{align}
f(0) = 1 \,,\qquad
f(+\infty) = 0 \,,\qquad
\int_0^\infty \dd u\s f(u)\,
&\quad\text{well-defined}\,, \\[6pt]
\left. u^n \frac{\dd^n f}{\dd u^n} \right|_{u=0} = \left. u^n \frac{\dd^n f}{\dd u^n} \right|_{u\to+\infty} = \,0
&\quad\text{for}\quad
n\ge1 \,.
\end{align}
\end{subequations}
Typical examples of such functions are
\begin{equation}
f(u) = e^{-u} \,,\qquad \text{and} \qquad
f(u) = \frac{2}{(1+u)(2+u)} \,.
\end{equation}
The \emph{renormalized} anomaly is then given by
\begin{equation}
\Acal_\beta [\alpha] \equiv \lim\limits_{\Lambda\to\infty} \Bigl(\Acal_\beta^\Lambda [\alpha] + \delta_\alpha \int \dd^4x\, \L_\text{ct}^\Lambda\Bigr) \,,
\label{eqn:AcalRenormalized}
\end{equation}
where $\L_\text{ct}^\Lambda$ is the local counterterm Lagrangian. Note in particular that the regularized anomaly $\Acal_\beta^\Lambda [\alpha]$ generically contains an $\O(\Lambda^2)$ piece that is irrelevant for $\beta$ values satisfying the Wess-Zumino consistency condition, in which case we should include operators with appropriate $\O(\Lambda^2)$ coefficients in $\L_\text{ct}^\Lambda$ to obtain a finite result for the renormalized anomaly $\Acal_\beta[\alpha]$. $\L_\text{ct}^\Lambda$ can also contain $\O(\Lambda^0)$ counterterms, and their coefficients specify the renormalization scheme.

Having included the damping factor $f\big(\!-\! \Pb{2} / \Lambda^2 \big)$, the functional trace $\Acal_\beta^\Lambda[\alpha]$ is now the sum of an absolutely convergent series. So at this point one is free to manipulate this expression, \eg\ perform cyclic permutations while maintaining a genuine `$=$' sign. Our regularization prescription \cref{eqn:AcalRegularized} is designed to facilitate the evaluation with CDE. In particular, the damping factor inserted commutes with the $\beta$-modified covariant derivative:
\begin{equation}
\comm{f\left(-\frac{\Pb{2}}{\Lambda^2}\right)}{\Pb{}} = 0 \,.
\end{equation}
Also note from the definition of $\Pb{}$ in \cref{eqn:PslbetaDef} that it anticommutes with $\gamma^5$:
\begin{equation}
\Pb{} \gamma^5 = -\gamma^5 \Pb{} \,.
\end{equation}
Making use of these identities, we can simplify \cref{eqn:AcalRegularized} to
\begin{align}
\Acal_\beta^\Lambda [\alpha] = \Tr \left[ f \left( -\frac{\Pb{2}}{\Lambda^2} \right) \alpha\, \gamma^5 \right] \,,
\label{eqn:AcalRegularizedSimplified}
\end{align}
from which it is clear that the evaluation result will depend on the parameters $\beta$. 

One interpretation of this regulator is that the $\beta$ parameters determine the combination of background fields that are turned on when computing the anomaly.  This effectively forces the path integral measure $\Dcal\chi \s \Dcal\chi^\dagger$ to be organized according to the eigenmodes of the operator $\Pb{}$.\footnote{We leave implicit possible analytic continuations needed to make $\Pb{}$ a Hermitian operator that has a well-defined eigenvalue problem.}

We will proceed with the evaluation of $\Acal_\beta^\Lambda[\alpha]$ in \cref{sec:Evaluation}, after the next two subsections which discuss how our prescription connects to other familiar regularization approaches, and how the Wess-Zumino consistency condition is satisfied or violated by different choices of $\beta$.

\subsection{Connection to Other Regularization Prescriptions}
\label{subsec:Connections}

Let us make a few comments on the connection between our regularization prescription \cref{eqn:AcalRegularized} and some approaches that often appear in the literature, in particular,  heat kernel regularization and Pauli-Villars regularization.

Regularizing \cref{eqn:Tbeta} with a heat kernel regulator, one obtains
\begin{equation}
\Acal_\beta^\text{HK} \equiv \Tr \left[ e^{\Pb{2}/\Lambda^2} \frac{1}{\Pb{}} \left( \alpha\, \Pb{} - \Pb{}\, \alpha \right) \frac{1-\gamma^5}{2} \right] \,.
\label{eqn:TbetaHK}
\end{equation} 
Comparing with \cref{eqn:AcalRegularized}, we see that the heat kernel regularization amounts to choosing the damping function to be
\begin{equation}
\text{Heat kernel}:\qquad
f(u) = e^{-u} \,.
\end{equation}

Alternatively, regularizing \cref{eqn:Tbeta} with one Pauli-Villars field, one obtains
\begin{align}
\Acal_\beta^\text{PV,1} &\equiv \Tr \left[ \left( \frac{1}{\Pb{}} - \frac{1}{\Pb{}-\Lambda} \right) \left( \alpha\, \Pb{} - \Pb{}\, \alpha \right) \frac{1-\gamma^5}{2} \right]
\notag\\[5pt]
&= \Tr \left[ \frac{-\Lambda}{\Pb{} (\Pb{} - \Lambda)} \left( \alpha\, \Pb{} - \Pb{}\, \alpha \right) \frac{1-\gamma^5}{2} \right] = \Tr \left[ \frac{-\Lambda}{\Pb{} - \Lambda}\, \alpha\, \gamma^5 \right]
\notag\\[5pt]
&= \Tr \left[ \frac{-\Lambda^2}{\Pb{2} - \Lambda^2}\, \alpha\, \gamma^5 \right]
= \Tr \left[ \frac{-\Lambda^2}{\Pb{2} - \Lambda^2} \frac{1}{\Pb{}} \left( \alpha\, \Pb{} - \Pb{}\, \alpha \right) \frac{1-\gamma^5}{2} \right] \,.
\label{eqn:TbetaPV1}
\end{align}
Comparing with \cref{eqn:AcalRegularized}, we see that this amounts to choosing the damping function to be
\begin{equation}
\text{Pauli-Villars with one regulator field}:\qquad
f(u) = \frac{1}{1+u} \,.
\end{equation}
Note that this damping factor does not satisfy all the conditions listed in \cref{eqn:fconditions}, and hence would not regulate all the divergences. This motivates considering Pauli-Villars regularization with three regulator fields, for which one obtains the regularized anomaly as
\begin{align}
\Acal_\beta^\text{PV,3} &\equiv \Tr \left[ \left( \frac{1}{\Pb{}} - \frac{1}{\Pb{}-M_1} + \frac{1}{\Pb{}-M_2} - \frac{1}{\Pb{}-M_3} \right) \left( \alpha\, \Pb{} - \Pb{}\, \alpha \right) \frac{1-\gamma^5}{2} \right]
\notag\\[5pt]
&= \Tr \left[ \frac{-(M_1-M_2+M_3)\Pb{2} + 2M_1M_3 \Pb{} - M_1M_2M_3}{(\Pb{}-M_1)(\Pb{}-M_2)(\Pb{}-M_3)}\, \alpha\, \gamma^5 \right]
\notag\\[5pt]
&= \Tr \left[ \frac{M_1^2 M_3^2\, (2\Pb{2}-M_2^2)}{(\Pb{2}-M_1^2)(\Pb{2}-M_2^2)(\Pb{2}-M_3^2)}\, \alpha\, \gamma^5 \right] \,,
\label{eqn:TbetaPV3}
\end{align}
where we have assumed the relation $M_1^2 - M_2^2 + M_3^2 = 0$. If we now take
\begin{equation}
M_1^2=M_3^2=\Lambda^2 \,,\qquad\text{and}\qquad
M_2^2 = 2\Lambda^2 \,,
\end{equation}
this simplifies to
\begin{equation}
\Acal_\beta^\text{PV,3} = \Tr \left[ \frac{2\Lambda^4}{(\Pb{2} - \Lambda^2)(\Pb{2} - 2\Lambda^2)} \frac{1}{\Pb{}} \left( \alpha\, \Pb{} - \Pb{}\, \alpha \right) \frac{1-\gamma^5}{2} \right] \,.
\label{eqn:TbetaPV3Lambda}
\end{equation}
Comparing with \cref{eqn:AcalRegularized}, we see that this amounts to choosing the damping function to be
\begin{equation}
\text{Pauli-Villars with three regulator fields}:\qquad
f(u) = \frac{2}{(1+u)(2+u)} \,.
\end{equation}
This damping function does satisfy all the conditions listed in \cref{eqn:fconditions}, and will successfully regularize all the divergences.

\subsection{Consistency With the Wess-Zumino Condition}
\label{subsec:WessZumino} 

Since we have adopted the definition of anomaly as the gauge variation of the bosonic effective action:
\begin{equation}
\Acal[\alpha] \equiv \delta_\alpha W[G]= (W[G_\alpha] - W[G])|_{\O(\alpha)} \,,
\end{equation}
we expect it to satisfy the Wess-Zumino consistency condition, as reviewed in \cref{subsec:Def}. However, an implicit assumption behind this expectation is that there is a well-defined $W[G]$. Importantly, our regularization prescription presented in \cref{subsec:beta} is directly applied to $\delta_\alpha W[G]$, instead of $W[G]$. In this case, the Wess-Zumino consistency condition may not be satisfied, because generic $\beta$ values may not correspond to applying the same (or `consistent') regularization prescription to $W[G_\alpha]$ and $W[G]$. In this subsection, we check the Wess-Zumino consistency condition for the regularized anomaly $\Acal_\beta^\Lambda [\alpha]$ at the level of \cref{eqn:AcalRegularizedSimplified}, and give a partial but general answer to the question of what $\beta$ values lead to a Wess-Zumino consistent anomaly:
\begin{equation}
\delta_{\alpha_1} \Acal_\beta^\Lambda [\alpha_2] - \delta_{\alpha_2} \Acal_\beta^\Lambda [\alpha_1] \stackrel{?}{=}  \Acal_\beta^\Lambda \bigl[-i\comm{\alpha_1}{\alpha_2}\bigr] \,.
\label{eqn:WZquestion}
\end{equation}
We will revisit this question in \cref{sec:Application} after evaluating \cref{eqn:AcalRegularizedSimplified} in \cref{sec:Evaluation}.

Using the expression of $\Acal_\beta^\Lambda [\alpha]$ in \cref{eqn:AcalRegularizedSimplified}, we can write the first term in \cref{eqn:WZquestion} as (it is understood that we will be dropping terms of order $\O(\alpha_1^2, \alpha_2^2)$ throughout this subsection):
\begin{equation}
\delta_{\alpha_1} \Acal_\beta^\Lambda [\alpha_2] = \Tr \left[ f\left( -\frac{\Pb{2}[\alpha_1]}{\Lambda^2} \right) \gamma^5 \alpha_2 \right] - \Tr \left[ f\left( -\frac{\Pb{2}}{\Lambda^2} \right) \gamma^5 \alpha_2 \right] \,,
\label{eqn:delta12}
\end{equation}
where $\Pb{}[\alpha_1]$ denotes the gauge transformation of $\Pb{}$:
\begin{align}
\Pb{}[\alpha_1] &\equiv i\pdsl + \Gsl_{\alpha_1} \left( \frac{1-\gamma^5}{2} + \beta \frac{1+\gamma^5}{2} \right) \notag\\[5pt]
&= i\pdsl + \Big[ U_{\alpha_1} \Gsl U_{\alpha_1}^\dagger + U_{\alpha_1} \left( i\pdsl U_{\alpha_1}^\dagger \right) \Big] \left( \frac{1-\gamma^5}{2} + \beta \frac{1+\gamma^5}{2} \right) \,.
\label{eqn:PbalphaDef}
\end{align}

We note that when $\beta=1$, \cref{eqn:delta12} is quite easy to calculate because $\widehat{P}_{\beta=1}=\Psl$ transforms covariantly and so does the damping factor:
\begin{equation}
\widehat{P}_{\beta=1}[\alpha_1] = U_{\alpha_1} \widehat{P}_{\beta=1} U_{\alpha_1}^\dagger
\quad\Longrightarrow\quad
f\left( -\frac{\widehat{P}_{\beta=1}^2[\alpha_1]}{\Lambda^2} \right) = U_{\alpha_1} f\left( -\frac{\widehat{P}_{\beta=1}^2}{\Lambda^2} \right) U_{\alpha_1}^\dagger \,.
\label{eqn:PbCovariant}
\end{equation}
This leads us to the so-called covariant anomaly that satisfies
\begin{equation}
\delta_{\alpha_1} \Acal_{\beta=1}^\Lambda [\alpha_2]
= \Tr \left[ f\left( -\frac{\widehat{P}_{\beta=1}^2}{\Lambda^2} \right) \gamma^5 \left( U_{\alpha_1}^\dagger\, \alpha_2\, U_{\alpha_1} - \alpha_2 \right) \right]
= \Acal_{\beta=1}^\Lambda \bigl[-i\comm{\alpha_1}{\alpha_2} \bigr] \,.
\end{equation}
We see that this covariant anomaly generically would not satisfy the Wess-Zumino consistency condition; it is off by a factor of two compared to \cref{eqn:WZquestion}:
\begin{equation}
\delta_{\alpha_1} \Acal_{\beta=1}^\Lambda [\alpha_2] - \delta_{\alpha_2} \Acal_{\beta=1}^\Lambda [\alpha_1] = 2\, \Acal_{\beta=1}^\Lambda \bigl[-i\comm{\alpha_1}{\alpha_2} \bigr] \quad\ne\quad \Acal_{\beta=1}^\Lambda \bigl[-i\comm{\alpha_1}{\alpha_2} \bigr] \,.
\end{equation}
The only exceptions are when the anomaly itself vanishes $\Acal_{\beta=1}^\Lambda [\alpha]=0$ (once summed over fermion species) or when the two gauge transformations under consideration commute, $\comm{\alpha_1}{\alpha_2}=0$. In these cases, the Wess-Zumino consistency condition itself is trivial, and the covariant anomaly is also a consistent anomaly.

For general $\beta$ values, $\Pb{}$ does not transform covariantly, and calculating \cref{eqn:delta12} is more tedious. It is useful to write $\Pb{}$ in terms of its chirality components:
\begin{equation}
\Pb{} = \Psl \frac{1-\gamma^5}{2} + \Psl_\beta \frac{1+\gamma^5}{2} \,,
\label{eqn:Pbdecomposition}
\end{equation}
with
\begin{subequations}\label{eqn:Pbcomponents}
\begin{align}
\Psl &= i\slashed{\partial} + \Gsl = i\slashed{\partial} +\sum_a\Gsl^a t^a \,, \\[5pt]
\Psl_\beta &\equiv i\slashed{\partial} + \beta\Gsl = i\slashed{\partial} +\sum_a\beta_a\Gsl^a t^a \,.
\end{align}
\end{subequations}
The left-handed component is gauge covariant, but the right-handed component transforms in a complicated manner for general $\beta$ values. To proceed, let us rewrite \cref{eqn:PbalphaDef} also in terms of its chirality components:
\begin{equation}
\Pb{} \quad\longrightarrow\quad
\Pb{} [\alpha_1] = \Lsl_{\alpha_1} \frac{1-\gamma^5}{2} + \Rsl_{\alpha_1} \frac{1+\gamma^5}{2} \,,
\end{equation}
with
\begin{subequations}
\begin{align}
\Psl \quad\longrightarrow\quad
\Lsl_{\alpha_1} &\equiv U_{\alpha_1} \Psl U_{\alpha_1}^\dagger \,, \\[5pt]
\Psl_\beta \quad\longrightarrow\quad
\Rsl_{\alpha_1} &\equiv i\pdsl + \beta \Big[ U_{\alpha_1} \Gsl U_{\alpha_1}^\dagger + U_{\alpha_1} \left( i\pdsl U_{\alpha_1}^\dagger \right) \Big] \,.
\end{align}
\end{subequations}
To check the Wess-Zumino consistency condition \cref{eqn:WZquestion}, we can Taylor expand the damping factors in \cref{eqn:delta12} and examine a general $k^\text{th}$ power term therein. We have
\begin{subequations}
\begin{align}
\Tr \bigg[ \left( \Pb{2}[\alpha_1] \right)^k \gamma^5 \alpha_2 \bigg]
&= \Tr \bigg\{ \bigg[ \left( \Rsl_{\alpha_1} \Lsl_{\alpha_1} \right)^k \tfrac{1-\gamma^5}{2} + \left( \Lsl_{\alpha_1} \Rsl_{\alpha_1} \right)^k \tfrac{1+\gamma^5}{2} \bigg] \gamma^5 \alpha_2 \bigg\}
\notag\\[5pt]
&= \Tr \bigg\{ \tfrac{1+\gamma^5}{2} \left( \Psl U_{\alpha_1}^\dagger \Rsl_{\alpha_1} U_{\alpha_1} \right)^{k-1} \Psl\, U_{ \alpha_1}^\dagger \comm{\Rsl_{\alpha_1}}{\alpha_2} U_{\alpha_1} \bigg\} \,,\label{eqn:Tr_alpha1}\\[10pt]
\Tr \bigg[ \left( \Pb{2} \right)^k \gamma^5 \alpha_2 \bigg]
&= \Tr \bigg\{ \tfrac{1+\gamma^5}{2} \left( \Psl \Psl_\beta \right)^{k-1} \Psl \comm{\Psl_\beta}{\alpha_2} \bigg\} \,.\label{eqn:Tr_0}
\end{align}
\end{subequations}
The difference between \cref{eqn:Tr_alpha1,eqn:Tr_0} comes from two sources:
\begin{subequations}\label{eqn:sources}
\begin{align}
U_{ \alpha_1}^\dagger \Rsl_{\alpha_1} U_{\alpha_1} &= \Psl_\beta + (\Rsl_{\alpha_1} - \Psl_\beta) + i\comm{\Psl_\beta}{\alpha_1} \,, \\[5pt]
U_{ \alpha_1}^\dagger \comm{\Rsl_{\alpha_1}}{\alpha_2} U_{\alpha_1} &= \comm{\Psl_\beta}{\alpha_2} + \comm{\Rsl_{\alpha_1}-\Psl_\beta}{\alpha_2} -i \Bigl[ \alpha_1, \comm{\Psl_\beta}{\alpha_2} \Bigr] \,.
\end{align}
\end{subequations}
One could go ahead with the calculation keeping track of all these terms for general $\beta$ values, but the result is not very illuminating. Instead, let us examine the special case $\beta=0$ here. In this case, the right-handed component does not transform:
\begin{equation}
\Rsl_{\alpha_1} = \Psl_{\beta=0} = i\pdsl \,.
\end{equation}
The middle term of each equation in Eqs.~(\ref{eqn:sources}) is therefore absent, and we have
\begin{align}
\delta_{\alpha_1} \Acal_{\beta=0}^\Lambda [\alpha_2] - \delta_{\alpha_2} \Acal_{\beta=0}^\Lambda [\alpha_1] &\supset
\Tr \bigg[ \left( \Pb{2}[\alpha_1] \right)^k \gamma^5 \alpha_2 \bigg] - \Tr \bigg[ \left( \Pb{2} \right)^k \gamma^5 \alpha_2 \bigg] - \left( \alpha_1 \leftrightarrow \alpha_2 \right)
\notag\\[6pt]
&= \Tr \bigg\{ \tfrac{1+\gamma^5}{2} \Big[ \left( \Psl U_{\alpha_1}^\dagger \Rsl_{\alpha_1} U_{\alpha_1} \right)^{k-1} - \left( \Psl \Psl_\beta \right)^{k-1} \Big] \Psl \comm{\Psl_\beta}{\alpha_2}
\notag\\[2pt]
&\hspace{30pt}
+ \tfrac{1+\gamma^5}{2} \left( \Psl \Psl_\beta \right)^{k-1} \Psl \Bigl[ -i\alpha_1, \comm{\Psl_\beta}{\alpha_2} \Bigr] \bigg\} - \left( \alpha_1 \leftrightarrow \alpha_2 \right) 
\notag\\[6pt]
&= \Tr \bigg\{ \tfrac{1+\gamma^5}{2} \left( \Psl \Psl_\beta \right)^{k-1} \Psl \Bigr[ \Psl_\beta, -i\,\comm{\alpha_1}{\alpha_2} \Bigr] \bigg\}
\notag\\[6pt]
&= \Tr \bigg[ \left( \Pb{2} \right)^k \gamma^5\, \comm{-i\alpha_1}{\alpha_2} \bigg] \,.
\label{eqn:kth}
\end{align}
In the step leading to the second to last line, the first term in the curly brackets gets canceled upon adding the expression with $\alpha_1 \leftrightarrow \alpha_2$, while the second term combines with the latter and we have used the Jacobi identity. Clearly, summing over all the $k^\text{th}$ power relations like in \cref{eqn:kth} will give us the Wess-Zumino consistency condition in \cref{eqn:WZquestion}:
\begin{equation}
\delta_{\alpha_1} \Acal_{\beta=0}^\Lambda [\alpha_2] - \delta_{\alpha_2} \Acal_{\beta=0}^\Lambda [\alpha_1]
= \Acal_{\beta=0}^\Lambda \bigl[-i\comm{\alpha_1}{\alpha_2}\bigr] \,.
\end{equation}
Therefore, we see that in our regularization prescription, $\beta=0$ (meaning $\beta_a=0$, $\forall a$) is always one possible choice to ensure the Wess-Zumino consistency condition for any symmetry group. However, from the present analysis it is difficult to tell whether there are other Wess-Zumino consistent choices. We will revisit this issue in \cref{sec:Application} using the BRST version of the Wess-Zumino condition once we have the evaluation result for $\Acal_\beta^\Lambda[\alpha]$.

\section{Master Formula for the Anomaly From CDE}
\label{sec:Evaluation}

Now we proceed with the evaluation of the regularized anomaly, starting with \cref{eqn:AcalRegularizedSimplified}:
\begin{align}
\Acal_\beta^\Lambda [\alpha] &= \Tr \Biggl[ f \Biggl( -\frac{\Pb{2}}{\Lambda^2} \Biggr) \alpha\, \gamma^5 \Biggr]
= \int \dd^4 x \int \frac{\dd^4 q}{(2\pi)^4} \tr \Biggl\{ f\Biggl[ - \frac{ \bigl(\qsl + \Pb{}\bigr)^2 }{\Lambda^2} \Biggr] \alpha\, \gamma^5 \Biggr\}
\notag\\[8pt]
&= \int \dd^4 x \int \frac{\dd^4 k}{(2\pi)^4} \tr \Biggl\{ \Lambda^4 f\Biggl[ - \biggl(\ksl + \frac{\Pb{}}{\Lambda} \biggr)^2 \Biggr] \alpha\, \gamma^5 \Biggr\} \,.
\label{eqn:TbetaLambdak}
\end{align}
Here we have rescaled the integration variable $k^\mu \equiv q^\mu/\Lambda$, so that it is easier to keep track of the $1/\Lambda$ powers. Eventually, we are interested in the $\Lambda\to\infty$ limit, so in what follows we will be dropping the $\O(1/\Lambda)$ terms that vanish in this limit. This can be achieved by applying the simplified CDE, while expanding and truncating the integrand accordingly:
\begin{align}
\Lambda^4 f\left[ - \left(\ksl + \frac{\Pb{}}{\Lambda} \right)^2 \right] &= \Lambda^4 f \left[ -k^2 - \frac{1}{\Lambda} \left( \ksl \Pb{} + \Pb{} \ksl \right) - \frac{1}{\Lambda^2} \Pb{2} \right]
\notag\\[5pt]
&\hspace{0pt}
= \Lambda^4 \left( f_u + f'_u\, z + \frac12 f''_u\, z^2 + \frac16 f'''_u\, z^3 + \frac{1}{24} f''''_u\, z^4 \right)\,,
\end{align}
where we have introduced the following notation for convenience
\begin{equation}
u \equiv -k^2 \,,
\qquad\text{and}\qquad
z \equiv - \frac{1}{\Lambda} \left( \ksl\Pb{} + \Pb{}\ksl \right) - \frac{1}{\Lambda^2} \Pb{2} \,.
\end{equation}
Plugging this back into \cref{eqn:TbetaLambdak} and simplifying the expression, we get
\begin{align}
\Acal_\beta^\Lambda [\alpha] &= \int \dd^4 x \int \frac{\dd^4 k}{(2\pi)^4} \tr \Bigg\{ \bigg[
- \Lambda^2 f'_u\, \Pb{2}
+ \frac12 f''_u\, \Pb{4}
\notag\\[5pt]
&\hspace{120pt}
+ \frac{u}{24} f'''_u \Big( \Pb{2} \gamma_\mu\Pb{} \gamma^\mu \Pb{}
+ \Pb{} \gamma_\mu\Pb{} \gamma^\mu \Pb{2}
\notag\\
&\hspace{170pt}
+ \Pb{} \gamma_\mu\Pb{2} \gamma^\mu \Pb{}
+ 4 \Pb{4} \Big) \bigg] \alpha\, \gamma^5 \Bigg\} \,.
\end{align}
Note that the terms proportional to $f_u''''$ can be grouped in pairs that take the form $\tr\bigl[\gamma^\mu (\cdots) \gamma^5 + (\cdots) \gamma^\mu \gamma^5\bigr] = 0$, so they all cancel out; the same is true for a subset of the $f_u''$ and $f_u'''$ terms, which significantly reduces the number of terms in the result.
Performing the loop momentum integral (after a Wick rotation as usual), we obtain
\begin{align}
\Acal_\beta^\Lambda [\alpha] &= \int \dd^4x\, \frac{i}{16\pi^2} \bigg\{ 
-\Lambda^2 \left[ \left( u f_u \right)\bigr|_0^\infty - \int_0^\infty\dd u\s f_u  \right] \tr_0
\notag\\[8pt]
&\hspace{80pt}
+ \frac12 \Big[ \left( u f'_u - f_u \right)\bigr|_0^\infty \Big] \tr_1
\notag\\[8pt]
&\hspace{80pt}
+ \frac{1}{12} \Big[ \left( u^2 f''_u - 2 u f'_u + 2f_u \right)\bigr|_0^\infty \Big] \left( 2 \tr_1 - \tr_2 - \tr_3 \right) \bigg\} \,.
\end{align}
We see that for a general damping function $f(u)$ that satisfies the conditions in Eqs.~(\ref{eqn:fconditions}), the calculation yields the result:
\begin{align}
\Acal_\beta^\Lambda [\alpha] &=\int \dd^4x\, \frac{i}{16\pi^2}\, \Bigg\{ \left[ \Lambda^2 \int_0^\infty\dd u\s f(u)  \right] \tr_0 + \frac{1}{6} \bigl(\tr_1 + \tr_2 + \tr_3\bigr) \Bigg\} \,,
\label{eqn:Master0}
\end{align}
where
\begin{subequations}
\label{eqn:DiracTraces}
\begin{align}
\tr_0 &\equiv \tr \Big[ \Pb{2} \gamma^5 \alpha \Big] \,,\\[8pt]
\tr_1 &\equiv \tr \Big[ \Pb{4} \gamma^5 \alpha \Big] \,,\\[8pt]
\tr_2 &\equiv -\frac{1}{2} \tr \Big[ \big( \Pb{2} \gamma_\mu \Pb{} \gamma^\mu \Pb{} + \Pb{} \gamma_\mu \Pb{} \gamma^\mu \Pb{2} \bigr) \gamma^5 \alpha \Big] \,,\\[8pt]
\tr_3 &\equiv -\frac{1}{2} \tr \Big[ \Pb{} \gamma_\mu \Pb{2} \gamma^\mu \Pb{} \gamma^5 \alpha \Big] \,.
\end{align}
\end{subequations}
\cref{eqn:Master0} is our master formula for the regularized anomaly before evaluation of the Dirac traces.

\subsection{Evaluating the Dirac Traces}
\label{subsec:DiracTraces}

In order to evaluate the Dirac traces in \cref{eqn:DiracTraces}, it is convenient to use the chirality decomposition of $\Pb{}$ in \cref{eqn:Pbdecomposition}:
\begin{equation}
\Pb{} = \Psl \frac{1-\gamma^5}{2} + \Psl_\beta \frac{1+\gamma^5}{2} \,,
\end{equation}
where
\begin{subequations}
\begin{align}
\Psl &\equiv i\slashed{\partial} + \Gsl = i\slashed{\partial} +\sum_a\Gsl^a t^a \,, \\
\Psl_\beta &\equiv i\slashed{\partial} + \beta\Gsl = i\slashed{\partial} +\sum_a\beta_a\Gsl^a t^a \,. \label{eq:Pbeta}
\end{align}
\end{subequations}
We also introduce the notation
\begin{equation}
G_-^\mu \equiv P^\mu - P_\beta^\mu = (1-\beta)\, G^\mu = \sum_a (1-\beta_a)\, G^{a\mu}\, t^a \,.
\label{eq:Gminus}
\end{equation}
The evaluation of tr$_0$ is straightforward:
\begin{align}
\tr_0 &\,= 2 \tr \big( \comm{P_\mu}{P_\beta^\mu} \alpha \big)
= -2\s i\s (1-\beta)\, \tr \big[ (\partial^\mu G_\mu)\, \alpha \big] \notag\\[8pt]
&\overset{\text{IBP}}{=} 2\s i\s (1-\beta)\, \tr \big[ G_\mu (\partial^\mu \alpha) \big]
= 2\s i \sum_a \tr (t^a t^b) (1-\beta_a)\, G_\mu^a\, (\partial^\mu \alpha^b) \,.
\label{eqn:tr0result}
\end{align}
Turning to $\tr_1, \tr_2, \tr_3$, we first note that they can be written in the following form:\footnote{To arrive at these expressions, we have used cyclic permutation to move $P_\beta^\mu$ to the right in half of the terms. Generally this is illegal since `tr' is only over the internal space while $P_\beta^\mu$ contains $\partial^\mu$ which is a spacetime operator. However, such cyclic permutations are innocuous in CDE calculations of functional traces that arise from evaluating the path integral at one loop. In fact, they have been used in many previous functional matching calculations. We clarify this subtle point in \cref{appsec:perm}.}
\begin{subequations}
\begin{align}
\tr_1 &= \frac{1}{2} \tr \Bigl[ \Psl \Psl_\beta \Psl \bigl[\Psl_\beta, \alpha\bigr] (1+\gamma^5) \Bigr] \,, \\[8pt]
\tr_2 &= \frac{1}{2} \tr \Bigl[ \bigl( \Psl \Psl_\beta^2 + \Psl_\beta^2 \Psl + \Psl^3\bigr) \comm{\Psl_\beta}{\alpha} (1+\gamma^5) + \Psl_\beta \Psl \Psl_\beta \comm{\Psl_\beta}{\alpha} (1-\gamma^5) \Bigr] \,, \\[8pt]
\tr_3 &= -4 \tr \Bigl[ \bigl( P_\nu P_\mu P_\beta^\mu + P_\beta^\mu P_{\mu} P_\nu\bigr) \comm{P_\beta^\nu}{\alpha} \Bigr] \,,
\end{align}
\end{subequations}
where we have used $\gamma^\mu \gamma^\nu \gamma_\mu = -2\gamma^\nu$, $\gamma^\mu \gamma^\nu \gamma^\rho \gamma_\mu = 4\eta^{\nu\rho}$ to simplify the products of gamma matrices. Upon evaluating the Dirac traces we can combine terms in the sum of all three traces such that all $P_\beta^\mu$ factors appear in commutators:
\begin{align}
\sum_{i=1}^3 \tr_i &= \tr \Biggl\{ -2\, \bigg(
	\Bigl[ 3\comm{P_\beta^\mu}{P_\beta^\nu} + 2\comm{P_\beta^\mu}{G_-^\nu}
	- \comm{P_\beta^\nu}{G_-^\mu} + \comm{G_-^\mu}{G_-^\nu} \;,\; G_{-\mu} \Bigr] 
\notag\\[-3pt]
&\hspace{60pt}
	- \Bigl[ P_{\beta,\mu} \;,\; \comm{P_\beta^\mu}{G_-^\nu} - \comm{G_-^\mu}{G_-^\nu} \Bigr]
	- G_-^\mu G_-^\nu G_{-\mu} \bigg) \comm{P_{\beta,\nu}}{\alpha}
\notag\\[5pt]
&\hspace{35pt}
	-i\eps_{\mu\nu\rho\sigma} \bigg[ \Bigl(
	2\s G_-^\mu G_-^\nu G_-^\rho
	+ \Bigl\{ 3 \bigl[ P_\beta^\mu, P_\beta^\nu \bigr] + 2\bigl[P_\beta^\mu, G_-^\nu\bigr] \;,\; G_-^\rho \Bigr\}
	\Bigr) \bigl[P_\beta^\sigma, \alpha\bigr]
\notag\\[-3pt]
&\hspace{85pt}
	+ 3 \bigl[ P_\beta^\mu, P_\beta^\nu \bigr] \bigl[ P_\beta^\rho , P_\beta^\sigma \bigr]\, \alpha\,
	\bigg] \Biggr\} \,.
\label{eqn:tr123result}
\end{align}
Having $P_\beta^\mu$ in commutators is important because it contains the derivative $\partial^\mu$, which as a functional operator is understood to act on everything to its right. But when it appears in a commutator, its action is local (or `closed') on the object appearing in the commutator; for example:\footnote{The local nature of all derivative operators in the CDE is also the reason why the otherwise illegal cyclic permutation in the internal trace `tr' in intermediate steps actually leads to the correct result; see \cref{appsec:perm} for a detailed discussion.}
\begin{equation}
\bigl[\partial^\mu, G_\mu(x)\bigr]\, \phi(x) = \partial^\mu G_\mu(x)\, \phi(x) - G_\mu(x)\, \partial^\mu \phi(x) = \bigl(\partial^\mu G_\mu(x)\bigr)\, \phi(x) \,.
\end{equation}

\subsection{The Evaluated Master Formula}
\label{subsec:Master}

Gathering the results in \cref{eqn:tr0result,eqn:tr123result} and substituting in \cref{eq:Pbeta,eq:Gminus} for $P_\beta^\mu$ and $G_-^\mu$, we obtain our evaluated master formula for the regularized anomaly expressed in the matrix notation:
\begin{align}
\Acal_\beta^\Lambda [\alpha] &=\int \dd^4x\, \frac{1}{16\pi^2} \tr \Bigg\{
-2\s (1-\beta) \left[ \Lambda^2 \int_0^\infty \dd u\s f(u) \right]  G_\mu \left( \partial^\mu \alpha \right)
\notag\\[5pt]
&\hspace{-10pt}
	+\frac13 (1-\beta) \biggl(
	i \Bigl[ (1+4\beta)\, \left(\partial_\mu G_\nu \right)
	- (1+2\beta)\, \left(\partial_\nu G_\mu \right)
	-i (1+3\beta^2) \bigl[ G_\mu, G_\nu\bigr] \;,\; G^\mu \Bigr]
\notag\\
&\hspace{40pt}
	+ \left( \partial^2 G_\nu \right) + i (1-2\beta) \bigl[ \left( \partial^\mu G_\mu \right) , G_\nu \bigr]
	- G^\mu (1-\beta) G_\nu (1-\beta) G_\mu \biggr)
	\left( D_\beta^\nu \alpha \right)
\notag\\[5pt]
&\hspace{-10pt}
	- \frac12\, \eps_{\mu\nu\rho\sigma} \bigg(
	\frac13\, \Bigl\{ (1-\beta) G^\rho \;,\; 2(1+2\beta) \left( \partial^\mu G^\nu \right) - i (1+2\beta+3\beta^2) G^\mu G^\nu \Bigr\} \left( D_\beta^\sigma \alpha \right)
\notag\\
&\hspace{40pt}
	+ 4\Big[ \beta \left( \partial^\mu G^\nu \right) - i \beta^2 G^\mu G^\nu \Big]
	\Big[ \beta \left( \partial^\rho G^\sigma \right) - i \beta^2 G^\rho G^\sigma \Big] \alpha \bigg) \Bigg\} \,,
\label{eqn:Master1}
\end{align}
where 
\begin{equation}
\left( D_\beta^\mu \alpha \right) \equiv \left( \partial^\mu \alpha \right) - i \beta \comm{G^\mu}{\alpha} \,.
\end{equation}
In \cref{eqn:Master1} we have carefully kept the $\beta$ factors in appropriate places such that each of them is associated with the gauge field that immediately follows it.

Depending on the application, it is sometimes more convenient to write out the adjoint components of the master formula in \cref{eqn:Master1}, which gives
\begin{align}
\Acal_\beta^\Lambda [\alpha] &= \int \dd^4x\, \frac{1}{16\pi^2} \Bigg\{
-\sum_{a,b} \tr(t^a t^b)\, (1-\beta_a) \bigg[
\,2\,\bigg( \Lambda^2 \int_0^\infty \dd u\s f(u) \bigg)  G_\mu^a \left( \partial^\mu \alpha^b \right)
\notag\\[2pt]
&\hspace{15pt}
+ \frac13\, \bigg\{ 
f^{aef} \Big[ (1+4\beta_a) \left( \partial_\mu G_\nu^e \right) - (1+2\beta_a) \left( \partial_\nu G_\mu^e \right) 
+ \left( 1 + 3\beta_a^2 \right) f^{egh} G_\mu^g G_\nu^h \Big] G^{f\mu} 
\notag\\
&\hspace{90pt}
- \left( \partial^2 G_\nu^a \right) + (1-2\beta_a) f^{aef} \left( \partial^\mu G_\mu^e \right) G_\nu^f \bigg\} 
\bigl(\partial_\nu\alpha^b +\beta_b f^{bcd} G_\nu^c \alpha^d\bigr) \bigg]
\notag\\[10pt]
&\hspace{-20pt}
-\sum_{a,b,c,d} \tr (t^a t^b t^c t^d)\, \frac13\, (1-\beta_a) (1-\beta_b) (1-\beta_c)\, G_\mu^a G_\nu^b G^{c\mu} \bigl(\partial_\nu\alpha^d +\beta_d f^{def} G_\nu^e \alpha^f\bigr)
\notag\\[8pt]
&\hspace{-20pt}
-\hspace{3pt}\sum_{a,b,c}\hspace{3pt} \tr(\{t^a, t^b\}\, t^c)\, \frac{1}{4}\, \eps_{\mu\nu\rho\sigma} \bigg[ \beta_a \beta_b 
\bigl( F_\text{lin}^{a\mu\nu} + \beta_a f^{ade} G^{d\mu} G^{e\nu} \bigr)
\bigl( F_\text{lin}^{b\rho\sigma} + \beta_b f^{bfg}G^{f\rho} G^{g\sigma} \bigr) \alpha^c
\notag\\
&\hspace{15pt}
+\frac13 \left( 1-\beta_b \right) \Big( 2(1+2\beta_a) F_\text{lin}^{a\mu\nu} +(1+2\beta_a+3\beta_a^2) f^{ade} G^{d\mu} G^{e\nu}\Big) G^{b\rho} 
\notag\\
&\hspace{90pt}
\times \bigl(\partial^\sigma\alpha^c +\beta_c f^{cfg} G^{f\sigma} \alpha^g\bigr) \bigg]
\Bigg\} \,,
\label{eqn:Master2}
\end{align}
where $F_\text{lin}^{\mu\nu} \equiv \left( \partial^\mu G^\nu \right) - \left( \partial^\nu G^\mu \right)$ is the part of $F^{\mu\nu}$ linear in the gauge fields, and we have used the fact that $\beta$ takes the same value within a simple group (only for which $f^{abc}$ may be nonzero).

In the next section, we will apply the evaluated master formula, written in matrix and component forms in
\cref{eqn:Master1,eqn:Master2}, respectively, to obtain explicit results for various gauge group sectors. Before delving into the details, let us first quickly note two special $\beta$ choices which directly relate to the discussion in \cref{subsec:WessZumino}.
\begin{itemize}
\item If $\beta_a=1$ ($\forall a$), all but the last line in \cref{eqn:Master1} vanishes, and the result takes a gauge-covariant form:
\begin{align}
\Acal_{\beta=1}^\Lambda [\alpha] &= \int \dd^4x\, \biggl(-\frac{1}{32\pi^2}\biggr)\,
\eps_{\mu\nu\rho\sigma}\, \tr \left( F^{\mu\nu} F^{\rho\sigma} \alpha \right) \notag\\[5pt]
&= \int \dd^4x\, \biggl(-\frac{1}{64\pi^2}\biggr)\, \tr(\{t^a, t^b\}\, t^c)\,
\eps_{\mu\nu\rho\sigma}\, F^{a\mu\nu} F^{b\rho\sigma} \alpha^c \,.
\end{align}
As discussed in \cref{subsec:WessZumino}, the covariant anomaly generically would not satisfy the Wess-Zumino consistency condition. However, we also mentioned some exceptions to this, such as when the anomaly itself is zero. From the equation above, we see that this can be achieved by the standard anomaly cancellation condition $\tr\,\bigl(\{t^a,t^b\}\,t^c\bigr) = 0$, where we recall that the internal trace `$\tr$' also sums over the fermion species.
\item If $\beta_a=0$ ($\forall a$), we learned from \cref{subsec:WessZumino} that the Wess-Zumino consistency condition should be satisfied. In this case, \cref{eqn:Master1} indeed reproduces the familiar result for the consistent anomaly:
\begin{align}
\Acal_{\beta=0}^\Lambda [\alpha] &= \int \dd^4x\, \frac{1}{16\pi^2}\, \tr \bigg\{
-2 \left[ \Lambda^2 \int_0^\infty\dd u\s f(u) \right] G_\mu \left( \partial^\mu \alpha \right)
\notag\\[5pt]
&\hspace{40pt}
	+\frac13\, \Big[ \left( \partial^2 G_\nu \right)
		+ i \comm{\left( \partial^\mu G_\mu \right)}{G_\nu}
		+ i \comm{F_{\mu\nu}}{G^\mu} - G_\mu G_\nu G^\mu \Big] \left( \partial^\nu \alpha \right)
\notag\\[5pt]
&\hspace{40pt}
	-\frac16\, \eps_{\mu\nu\rho\sigma} 
		\Big\{ G^\rho \;,\; 2 \left(\partial^\mu G^\nu \right) - i G^\mu G^\nu \Big\}
		\left( \partial^\sigma \alpha \right)
	\bigg\}
\notag\\[10pt]
&\hspace{-30pt}
	= \int \dd^4x \left\{ \frac{1}{48\pi^2}\, \eps^{\mu\nu\rho\sigma} \tr \Bigl[
	\left( \partial_\mu \alpha \right) \left( G_\nu F_{\rho\sigma} + i\s G_\nu G_\rho G_\sigma \right) \Bigr]
	-\delta_\alpha \L_{\text{ct},0}^\Lambda \right\}
\notag\\[10pt]
&\hspace{-30pt}
	= \int \dd^4x\, \biggl\{ \frac{1}{48\pi^2} \tr \left( \acomm{t^a}{t^b}, t^c \right) \eps^{\mu\nu\rho\sigma}
	\left( \partial_\mu \alpha^a \right) \left[ \left( \partial_\nu G_\rho^b \right) + \frac14\, f^{bde} G_\nu^d G_\rho^e \right] G_\sigma^c
\notag\\[3pt]
&\hspace{40pt}
	-\delta_\alpha \L_{\text{ct},0}^\Lambda \biggr\} \,,
\label{eq:A_beta=0}
\end{align}
up to an irrelevant anomaly given by the gauge variation of the following local counterterm:
\begin{align}
\L_{\text{ct},0}^\Lambda &= \frac{1}{16\pi^2} \left[ \Lambda^2 \int_0^\infty \dd u\s f(u) \right] \tr \bigl(G^\mu G_\mu\bigr)
\notag\\[5pt]
&\hspace{20pt}
+ \frac{1}{96\pi^2} \tr \biggl[ \bigl(\partial^\mu G_\mu\bigr)^2
- 2 i\s F^{\mu\nu} G_\mu G_\nu + \frac12\, G^\mu G^\nu G_\mu G_\nu \biggr] \,.
\label{eq:Lct0}
\end{align}
The relevant anomaly in \cref{eq:A_beta=0} is proportional to $\tr\,\bigl(\{t^a,t^b\}\,t^c\bigr)$, which depends on the fermion content of the theory. The symmetries under consideration can be gauged when there is no relevant anomaly, that is, when the standard anomaly cancellation condition $\tr\,\bigl(\{t^a,t^b\}\,t^c\bigr) = 0$ is satisfied.
\end{itemize}

\section{Implications of the Master Formula}
\label{sec:Application}

In this section, we apply \cref{eqn:Master1,eqn:Master2} derived in the previous section, which are evaluation results of our master formula \cref{eqn:Master0}, to obtain explicit results for the anomaly in all possible combinations of the continuous group sectors. We consider in turn a simple non-Abelian group, semi-simple product of non-Abelian sectors, product of Abelian sectors, and finally the general case of product of non-Abelian and Abelian sectors. In each case, we aim to answer the following questions:
\begin{itemize}
\item What values of the regularization parameters $\beta$ are consistent with the Wess-Zumino condition?
\item For these Wess-Zumino consistent $\beta$ choices, what is the relevant anomaly, and what are the counterterms associated with the irrelevant anomaly?
\item What are the conditions for the relevant anomaly to vanish (in which case the symmetries under consideration can be gauged in the quantum theory)?
\end{itemize}

To investigate the first question, we use the BRST form of the Wess-Zumino consistency condition, which states that (recall the discussion around \cref{eqn:AcalBRST}) when the gauge variation parameter $\alpha$ is replaced by the ghost field $\omega$, the anomaly is BRST-closed:
\begin{equation}
\delta_\brst \Acal_\beta [\omega] = 0 \,,
\end{equation}
where $\Acal_\beta [\omega]$ is understood as the renormalized anomaly defined in \cref{eqn:AcalRenormalized}. Since the gauge variation of local counterterms is always BRST-closed due to the nil-potency of the BRST transformation, this requires the regularized anomaly is also BRST-closed:
\begin{equation}
\delta_\brst \Acal_\beta^\Lambda [\omega] = 0 \,.
\end{equation}
We will check this condition up to $\O(1/\Lambda)$ terms. To do so, it is useful to recall that under the BRST transformation:
\begin{subequations}\label{eqn:BRST}
\begin{align}
\delta_\brst G_\mu &= D_\mu \omega = \partial_\mu \omega -i \bigl[ G_\mu, \omega \bigr] \,, \\[4pt]
\delta_\brst F_{\mu\nu} &= -i \bigl[ F_{\mu\nu} , \omega \bigr] \,, \\[4pt]
\delta_\brst \omega &= i\omega^2 \,, \\[4pt]
\delta_\brst \bigl( \partial_\mu \omega -i\beta \bigl[ G_\mu, \omega\bigr]\bigr) &= i (1-\beta) \, \bigl\{ \omega , \partial_\mu \omega \bigr\} \,.
\end{align}
\end{subequations}

In answering the second and third questions, we will see how the well-known results for anomalies are recovered in our formalism with specific (Wess-Zumino consistent) $\beta$ choices. We will also see that for all the Wess-Zumino consistent anomalies, the standard anomaly cancellation condition $\tr\,\bigl(\{t^a,t^b\}\,t^c\bigr) = 0$ will guarantee that the relevant anomaly vanishes (which means the symmetries can be gauged).

\subsection{Simple Non-Abelian Group}
\label{subsec:simple_nonAbelian}

For a simple non-Abelian group, all the $\beta$ factors are degenerate, so we omit their adjoint indices and simply write all of them as $\beta$. We can first verify that the $\O(\Lambda^2)$ term in \cref{eqn:Master1} is BRST-closed:
\begin{align}
\delta_\brst \Acal_\beta^\Lambda[\omega]\bigr|_{\O(\Lambda^2)}
&= -\frac{1}{8\pi^2} \int \dd^4x\, \left[ \Lambda^2 \int_0^\infty \dd u\s f(u) \right] (1-\beta)
\notag\\[4pt]
&\hspace{80pt}
\times \tr \Big[ (\partial^\mu \omega) (\partial_\mu \omega) + i \big\{ \omega \,,\, G^\mu (\partial_\mu\omega) \big\} \Big] = 0 \,.
\end{align}
Note that cyclic permutation of a Grassmann odd matrix in the trace is accompanied by a minus sign if it passes through an odd number of Grassmann odd matrices, \eg\ $\tr\big[ \omega\, G^\mu (\partial_\mu\omega) \big] = -\tr\big[ G^\mu (\partial_\mu\omega)\, \omega \big]$.

To derive constraints on $\beta$ from the Wess-Zumino consistency condition, we need to consider the $\O(\Lambda^0)$ terms. The BRST transformation of these terms is quite tedious. However, as we will show, it turns out sufficient to work out just a subset of terms. Let us first note that $\Acal_\beta^\Lambda[\omega]|_{\O(\Lambda^0)}$ contains terms of the form:\vspace{-2pt}
\begin{equation}
\omega G   \partial^3 \,,\qquad
\omega G^2 \partial^2 \,,\qquad 
\omega G^3 \partial   \,,\qquad
\omega G^4\,,\vspace{-2pt}
\end{equation}
whose BRST transformation contains terms of the form:\footnote{Note that the $\omega^2 \partial^4$ term from BRST transforming the sole $\omega G \partial^3$ term in \cref{eqn:Master1} vanishes.}\vspace{-2pt}
\begin{equation}
\omega^2 G   \partial^3 \,,\qquad
\omega^2 G^2 \partial^2 \,,\qquad
\omega^2 G^3 \partial   \,,\qquad
\omega^2 G^4\,.\vspace{-2pt}
\end{equation}
We will see that the $\omega^2 G^4$ and $\omega^2 G \partial^3$ terms are sufficient to constrain $\beta$.

The $\omega^2 G^4$ terms can only come from BRST transforming the $\omega G^4$ terms in $\Acal_\beta^\Lambda [\omega]$. Those $\omega G^4$ terms that do not involve $\eps_{\mu\nu\rho\sigma}$ are easily seen to vanish upon cyclic permutation, and we are left with
\begin{align}
\Acal_\beta^\Lambda[\omega]\bigr|_{G^4\omega} &= 
\int \dd^4x\, \frac{1}{24\pi^2} \, \beta \,(1+\beta+\beta^2) \,\varepsilon_{\mu\nu\rho\sigma} \tr \bigl( G^\mu G^\nu G^\rho G^\sigma \omega \bigr)
\notag\\[8pt]
&= \int \dd^4x\, \biggl(-\frac{1}{192\pi^2}\biggr) \, \beta \,(1+\beta+\beta^2) \tr \bigl( \{ t^a , t^b \}\, t^c\bigr) 
\notag\\[5pt]
&\hspace{100pt}
\times \varepsilon_{\mu\nu\rho\sigma} f^{ade} f^{bfg} G^{d\mu} G^{e\nu} G^{f\rho} G^{g\sigma} \omega^c \,.
\end{align}
Since $\delta_\brst \Acal_\beta^\Lambda[\omega]\bigr|_{G^4 \omega^2} = 0$ requires $\Acal_\beta^\Lambda[\omega]\bigr|_{G^4 \omega} = 0$, while $(1+\beta+\beta^2)$ is positive-definite, we see that
\begin{equation}
\delta_\brst \Acal_\beta^\Lambda[\omega]\bigr|_{G^4 \omega^2} = 0
\qquad\Longrightarrow\qquad
\beta = 0 \quad \text{or} \quad \tr \bigl( \{ t^a , t^b \}\, t^c\bigr) = 0 \,.
\end{equation}
As discussed around \cref{eq:A_beta=0}, $\beta=0$ reproduces the standard consistent anomaly, plus an irrelevant piece that is obviously BRST-closed. The other option is the standard anomaly cancellation condition $\tr \bigl( \{ t^a , t^b \}\, t^c\bigr) = 0$; when this is true, the terms in $\Acal_\beta^\Lambda$ that are proportional to $\varepsilon_{\mu\nu\rho\sigma}$ all vanish. In this case, it remains to check whether there are additional constraints on the value of $\beta$ from the terms not involving $\varepsilon_{\mu\nu\rho\sigma}$. To do so, we focus on the $\omega^2 G \partial^3$ terms in $\delta_\brst \Acal_\beta^\Lambda[\omega]$, for which we find, after some simplification using cyclic permutation and integration by parts:
\begin{align}
\delta_\brst \Acal_\beta^\Lambda[\omega]\bigr|_{\omega^2 G \partial^3} &=
\int \dd^4x\, \frac{i}{48\pi^2} \, \beta\, (1-\beta) \tr \Bigl\{ \bigl( \partial^2 \bigl[G^\nu, \omega \bigr] - \bigl[ (\partial^2 G^\nu) , \omega\bigr]\bigr) (\partial_\nu \omega) \Bigr\}
\notag\\[8pt]
&= \int \dd^4x\, \frac{1}{48\pi^2} \, \beta \,(1-\beta) \tr (t^a t^b) 
\notag\\
&\hspace{90pt}
\times f^{acd} \Bigl[\bigl(\partial^2 G^{c\nu}\bigr) \omega^d - \partial^2 \bigl(G^{c\nu} \omega^d\bigr)\Bigr] (\partial_\nu \omega^b) \,.
\end{align}
Here the group theory factor $\tr(t^a t^b)\propto \delta^{ab}$ is always non-vanishing, so we see the only other option (besides $\beta=0$) that makes $\delta_\brst \Acal_\beta^\Lambda[\omega]\bigr|_{\omega^2 G \partial^3}$ vanish is $\beta=1$, in which case the $\eps_{\mu\nu\rho\sigma}$-independent part of $\Acal_\beta^\Lambda$ simply vanishes. 

In summary, we conclude that consistency with the Wess-Zumino condition requires either of the following to be true:
\begin{itemize}
\setlength\itemsep{3pt}
\item $\beta = 0$, in which case the result is given by \cref{eq:A_beta=0}. This reproduces the standard consistent anomaly plus an irrelevant piece that is equal to the gauge variation of the local counterterm given by \cref{eq:Lct0}. As discussed below \cref{eq:A_beta=0}, for the relevant anomaly to vanish in this case, one needs the standard anomaly cancellation condition $\tr\,\bigl(\{t^a,t^b\}\,t^c\bigr) = 0$.
\item $\tr\,\bigl(\{t^a,t^b\}\,t^c\bigr) = 0$ and $\beta =1$, in which case anomaly cancellation happens and the regularized anomaly vanishes altogether, $\Acal_\beta^\Lambda[\alpha]=0$.
\end{itemize}

\subsection{Product of Non-Abelian Sectors}
\label{subsec:nonAbelianProduct}

For a semi-simple product of non-Abelian sectors, the only additional term in $\Acal_\beta^\Lambda[\alpha]$ to consider is the $\tr(t^a t^b t^c t^d)$ term in \cref{eqn:Master2}. Both $\tr(t^a t^b)$ and $\tr(\{t^a, t^b\}\, t^c)$ vanish when the generators belonging to more than one simple sectors are involved, while $\tr(t^a t^b t^c t^d)$ can be nonzero when two of the four generators belong to one simple sector and the other two belong to another simple sector.

Upon imposing the conditions derived in the previous subsection on each simple non-Abelian sector, we see that there are only two scenarios. If $\beta=1$ (and $\tr(\{t^a, t^b\}\, t^c)=0$) for either sector, the aforementioned cross term in $\Acal_\beta^\Lambda[\alpha]$ vanishes because of the $(1-\beta_a)(1-\beta_b)(1-\beta_c)$ factor. If $\beta = 0$ for both sectors, the cross term is contained in the general result \cref{eq:A_beta=0}, specifically the gauge variation of the $\O(G^4)$ counterterm in \cref{eq:Lct0}. Therefore, no additional constraints arise from the Wess-Zumino consistency condition beyond those already derived for each simple sector. The same is true for the relevant anomaly cancellation condition.

\subsection{Product of Abelian Sectors}
\label{subsec:U1}

For an Abelian gauge group, we can set $f^{abc}=0$ and $F_\text{lin}^{\mu\nu} = F^{\mu\nu}$ in \cref{eqn:Master2} to obtain
\begin{align}
\Acal_\beta^\Lambda [\alpha] &= \int \dd^4x\, \frac{1}{16\pi^2} \bigg\{
\notag\\[3pt]
&\hspace{-10pt}
-\hspace{3pt}\sum_{a,b}\hspace{3pt} \tr(Q_a Q_b) \cdot (1-\beta_a) 
\biggl[
2 \left( \Lambda^2 \int_0^\infty\dd u\s f(u) \right)  G_\mu^a 
-\frac13 \left(\partial^2 G^a_\mu\right) \biggr] \bigl(\partial^\mu  \alpha^b\bigr)
\notag\\[5pt]
&\hspace{-10pt}
- \sum_{a,b,c,d} \tr (Q_a Q_b Q_c Q_d) \cdot \frac13\, (1-\beta_a) (1-\beta_b) (1-\beta_c)\, G^{a\mu} G^{b\nu} G_\mu^c \bigl(\partial_\nu\alpha^d \bigr)
\notag\\[5pt]
&\hspace{-10pt}
-\hspace{3pt}\sum_{a,b,c}\hspace{3pt} \tr(Q_a Q_b Q_c) \cdot
\frac{1}{8} \biggl[ (1+\beta_a)(1+\beta_b) +\frac{1}{3} (1-\beta_a)(1-\beta_b)\biggr]\,
\eps^{\mu\nu\rho\sigma} F^{a}_{\mu\nu} F^{b}_{\rho\sigma} \alpha^c\bigg\} \,,
\label{eq:A_beta_Abelian}
\end{align}
where we have written the group generators $t^a$ as $Q_a$ since they are just charges under the $U(1)$'s, and `tr' means summing over all chiral fermions. In the $\tr(Q_a Q_b Q_c)$ term, we have integrated by parts and symmetrized the coefficient between $a$ and $b$: 
\begin{equation}
\beta_a\beta_b +\frac{1}{3}(1+2\beta_a)(1-\beta_b)\to \frac{1}{4}\bigl[ (1+\beta_a)(1+\beta_b) +\frac{1}{3} (1-\beta_a)(1-\beta_b)\bigr]\,.
\end{equation}

Under the BRST transformation, only the gauge fields $G_\mu^a$ transform nontrivially while $F^a_{\mu\nu}$ and $\omega^a$ stay invariant, and we obtain
\begin{align}
\delta_\brst \Acal_\beta^\Lambda[\omega] =&\int \dd^4x\, \frac{1}{16\pi^2} \Bigg\{
\sum_{a,b} \tr(Q_a Q_b) \cdot (\beta_a-\beta_b) 
\notag\\
&\hspace{90pt}
\times \biggl[
\left( \Lambda^2 \int_0^\infty  \dd u\s f(u) \right) ( \partial_\mu \omega^a )
-\frac16 \left( \partial^2 \partial_\mu \omega^a \right)
\biggr](\partial^\mu  \omega^b)
\notag\\[10pt]
&\hspace{-20pt}
+\sum_{a,b,c,d} \tr (Q_a Q_b Q_c Q_d) \cdot \frac16\, (1-\beta_a) (1-\beta_b) (\beta_c-\beta_d)  
\notag\\[-5pt]
&\hspace{40pt}
\times\Bigl[ G^{a\mu} G_\mu^b \bigl(\partial^\nu\omega^c\bigr) \bigl(\partial_\nu\omega^d \bigr) + 2G^{a}_{\mu} G^{b}_{\nu} \bigl(\partial^\mu \omega^c\bigr) \bigl(\partial^\nu\omega^d \bigr) \Bigr]
\Bigg\} \,,
\label{eq:brst_Abelian}
\end{align}
where we have used the (anti-)symmetry between the adjoint indices to simplify the expression. From \cref{eq:brst_Abelian} we see that the Wess-Zumino consistency condition $\delta_\brst \Acal_\beta^\Lambda[\omega] = 0$ requires the following:
\begin{itemize}
\setlength\itemsep{6pt}
\item $\beta_a=\beta_b$ for any two Abelian sectors $a,b$ for which $\tr(Q_a Q_b)\ne0$. 
\item Either $\beta_a=\beta_b=\beta_c=\beta_d$ or at least two of them are equal to 1 for any group of Abelian sectors for which $\tr (Q_a Q_b Q_c Q_d)\ne0$.\footnote{This applies to groups of two, three and four Abelian sectors since $a,b,c,d$ do not have to be distinct.} 
\end{itemize}
When these conditions are satisfied, symmetrizing the indices allows one to show that the $\tr(Q_a Q_b)$ and $\tr(Q_a Q_b Q_c Q_d)$ terms in \cref{eq:A_beta_Abelian}, if nonzero, are equal to the gauge variation of local counterterms, and we have:
\begin{align}
\Acal_\beta^\Lambda [\alpha] = &\int \dd^4x\,  \biggl\{ -\delta_\alpha \L_\text{ct}^{(\beta)} -\frac{1}{128\pi^2} \sum_{a,b,c} \tr(Q_a Q_b Q_c) 
\nonumber\\[5pt]
&\hspace{40pt}
\times\biggl[ (1+\beta_a)(1+\beta_b) +\frac{1}{3} (1-\beta_a)(1-\beta_b)\biggr]\,
\varepsilon^{\mu\nu\rho\sigma} F^{a}_{\mu\nu} F^{b}_{\rho\sigma} \alpha^c
\biggr\} \,,
\label{eq:A_beta_Abelian_relevant}
\end{align}
where
\begin{align}
\L_\text{ct}^{(\beta)} &= \frac{1}{16\pi^2} \Bigg\{
\sum_{a,b}\tr(Q_a Q_b) (1-\beta_a) 
\biggl[
\left( \Lambda^2 \int_0^\infty \dd u\s f(u) \right)  G_\mu^a G^{b\mu}
-\frac{1}{6}\,\bigl(\partial^2 G^a_\mu \bigr) G^{b\mu} 
\biggr]
\notag\\[5pt]
&\hspace{15pt}
+ \sum_{a,b,c,d} \tr (Q_a Q_b Q_c Q_d) \cdot \frac{1}{12}\,  (1-\beta_a) (1-\beta_b) (1-\beta_c)  \,G^{a\mu} G^{b\nu} G_\mu^c G_\nu^d
\Bigg\} \,.
\label{eq:L_ct_Abelian}
\end{align}
Therefore, as in the non-Abelian case, a relevant anomaly may only come from terms with three gauge group generators. But unlike the non-Abelian case, $\beta$ values other than 0 and 1 are allowed. Again, we see that the standard anomaly cancellation condition $\tr\,\bigl(\{t^a,t^b\}\,t^c\bigr) = 0$ (\ie\ $\tr(Q_a Q_b Q_c)=0$ in the Abelian case) would guarantee that the relevant anomaly vanishes.\footnote{One may further ask whether the $\tr(Q_a Q_b Q_c)$ terms in \cref{eq:A_beta_Abelian_relevant} may also be irrelevant. Indeed, there are local counterterms of the form $\eps^{\mu\nu\rho\sigma}F^a_{\mu\nu} G^b_\rho G^c_\sigma$ one can write down. However, there may not be enough such counterterms to absorb all the anomalies; in particular, if $\tr(Q_a^3)\ne0$ for some Abelian sector $a$ there must be a relevant anomaly, since the counterterm above vanishes when $a=b=c$. \label{footnote:eps_ct}}

\subsubsection*{$U(1)_V\times U(1)_A$ Example}

Let us apply the results above to the classic example of two Abelian sectors $U(1)_V\times U(1)_A$. The matter content is assumed to consist of pairs of Weyl fermions with opposite (identical) charges under $U(1)_V$ ($U(1)_A$); the minimal case is that of two Weyl fermions with $(Q_V, Q_A) = (1, 1)$ and $(-1, 1)$, respectively. So the potentially nonzero traces are:
\begin{subequations}
\begin{align}
\tr(Q_V^2) \,,\quad \tr(Q_A^2) \,, \\[5pt]
\tr(Q_V^2 Q_A) \,,\quad \tr(Q_A^3) \,, \\[5pt]
\tr(Q_V^4) \,,\quad \tr(Q_V^2 Q_A^2) \,,\quad \tr(Q_A^4) \,.
\end{align}
\end{subequations}
The fact that $\tr(Q_V^2 Q_A^2)\ne0$ implies that to satisfy the Wess-Zumino consistency condition  we must choose
\begin{equation}
\beta_V = \beta_A \qquad\text{or}\qquad \beta_V=1 \qquad\text{or}\qquad \beta_A=1 \,.
\label{eq:WZ_U1VxU1A}
\end{equation}
Assuming one of these is true, we can readily obtain the anomaly result from \cref{eq:A_beta_Abelian_relevant}:
\begin{align}
\Acal_\beta^\Lambda [\alpha] = &\int \dd^4x\,  \biggl\{ -\delta_\alpha \L_\text{ct}^{(\beta_V^{},\beta_A^{})}
\nonumber\\[8pt]&
-\frac{1}{64\pi^2} \tr(Q_V^2 Q_A) \biggl[ (1+\beta_V)(1+\beta_A) +\frac{1}{3} (1-\beta_V)(1-\beta_A)\biggr]\,
\varepsilon_{\mu\nu\rho\sigma} F_V^{\mu\nu} F_A^{\rho\sigma} \alpha_V^{}
\nonumber\\[8pt]&
-\frac{1}{128\pi^2} \tr(Q_V^2 Q_A) \biggl[ (1+\beta_V)^2 +\frac{1}{3} (1-\beta_V)^2\biggr]\,
\varepsilon_{\mu\nu\rho\sigma} F_V^{\mu\nu} F_V^{\rho\sigma} \alpha_A^{}
\nonumber\\[8pt]&
-\frac{1}{128\pi^2} \tr(Q_A^3) \biggl[ (1+\beta_A)^2 +\frac{1}{3} (1-\beta_A)^2\biggr]\,
\varepsilon_{\mu\nu\rho\sigma} F_A^{\mu\nu} F_A^{\rho\sigma} \alpha_A^{}
\biggr\} \,.
\label{eq:A_beta_U1VxU1A}
\end{align}
As discussed in \cref{footnote:eps_ct}, there is in fact an additional possible counterterm, $\eps_{\mu\nu\rho\sigma}F_V^{\mu\nu} V^\rho A^\sigma$ (where $V$ and $A$ denote gauge fields), whose gauge variation produces a linear combination of $\eps_{\mu\nu\rho\sigma} F_V^{\mu\nu} F_A^{\rho\sigma} \alpha_V^{}$ and $\eps_{\mu\nu\rho\sigma} F_V^{\mu\nu} F_V^{\rho\sigma} \alpha_A^{}$ upon integration by parts. We will come back to this point shortly.

\cref{eq:A_beta_U1VxU1A} reproduces the standard result if we further demand that $U(1)_V$ is not anomalous and is preserved by renormalization. This means that we should pick the $\beta_V=1$ option in \cref{eq:WZ_U1VxU1A} so that $\L_\text{ct}^{(\beta_V^{},\beta_A^{})}$ does not involve $U(1)_V$-breaking operators (see \cref{eq:L_ct_Abelian}). This also rules out the additional counterterm $\eps_{\mu\nu\rho\sigma}F_V^{\mu\nu} V^\rho A^\sigma$ discussed above. For $U(1)_V$ to be non-anomalous, the coefficient of the $\alpha_V^{}$ term in \cref{eq:A_beta_U1VxU1A} must vanish, which requires $\beta_A=-1$ for $\beta_V=1$. We conclude that the standard result corresponds to the specific scheme choice in our formalism: 
\begin{equation}
(\beta_V\,,\, \beta_A) = (1\,,\, -1) \,,
\end{equation}
in which case \cref{eq:A_beta_U1VxU1A} becomes:
\begin{align}
\Acal_{(1,-1)}^\Lambda [\alpha] &= \int \dd^4x\,  \biggl\{ -\delta_\alpha \L_\text{ct}^{(1,-1)} 
\notag\\[8pt]
&\hspace{20pt}
-\frac{1}{32\pi^2}\,
\eps_{\mu\nu\rho\sigma} \biggl[\tr(Q_V^2 Q_A) \,F_V^{\mu\nu} F_V^{\rho\sigma} +\tr(Q_A^3)\cdot \frac13\, F_A^{\mu\nu} F_A^{\rho\sigma}\biggr] \alpha_A^{}
\biggr\} \,,
\label{eqn:U1VU1Abeta1m1}
\end{align}
with the following $U(1)_V$-preserving counterterm:
\begin{align}
\L_\text{ct}^{(1,-1)} =\frac{1}{16\pi^2} \biggl\{&
\tr(Q_A^2) 
\biggl[
2 \left( \Lambda^2 \int_0^\infty \dd u\s f(u) \right)  A^\mu A_\mu
-\frac13\, \bigl(\partial^2 A^\mu \bigr) A_\mu
\biggr]
\nonumber\\[8pt]&
+ \tr (Q_A^4) \cdot \frac{2}{3}\, (A^{\mu} A_\mu)^2 \biggr\} \,.
\end{align}

It is interesting to note that if we instead choose
\begin{equation}
(\beta_V\,,\, \beta_A) = (0\,,\, 0) \,,
\end{equation}
which is also Wess-Zumino consistent but does not manifestly preserve $U(1)_V$, we would obtain:
\begin{align}
\Acal_{(0,0)}^\Lambda [\alpha] &= \int \dd^4x\,  \biggl\{ -\delta_\alpha \L_\text{ct}^{(0,0)}
-\frac{1}{48\pi^2}\, \eps_{\mu\nu\rho\sigma} \tr(Q_V^2 Q_A) \,F_V^{\mu\nu} F_A^{\rho\sigma} \alpha_V^{}
\notag\\[8pt]
&\hspace{40pt}
-\frac{1}{96\pi^2}\,
\eps_{\mu\nu\rho\sigma} \biggl[\tr(Q_V^2 Q_A) \,F_V^{\mu\nu} F_V^{\rho\sigma} +\tr(Q_A^3)\, F_A^{\mu\nu} F_A^{\rho\sigma}\biggr] \alpha_A^{}
\biggr\} \,.
\end{align}
This is in fact related to the standard result \cref{eqn:U1VU1Abeta1m1} by a counterterm:
\begin{equation}
\Acal_{(0,0)}^\Lambda [\alpha] = \Acal_{(1,-1)}^\Lambda [\alpha] +\delta_\alpha \int \dd^4 x \,\bigl( \L_\text{ct}^{(1,1)} -\L_\text{ct}^{(0,0)} + \Delta\L_\text{ct} \bigr) \,,
\end{equation}
where
\begin{equation}
\Delta\L_\text{ct} = \frac{1}{24\pi^2}\,\eps_{\mu\nu\rho\sigma}\tr(Q_V^2 Q_A) \, F_V^{\mu\nu} V^\rho A^\sigma \,.
\end{equation}
Therefore, $(\beta_V,\beta_A)=(0,0)$ actually gives the same relevant anomaly as the standard result, although at the cost of $U(1)_V$-breaking counterterms. Note that it is impossible to remove both $\eps_{\mu\nu\rho\sigma} F_V^{\mu\nu} F_A^{\rho\sigma} \alpha_V^{}$ and $\eps_{\mu\nu\rho\sigma} F_V^{\mu\nu} F_V^{\rho\sigma} \alpha_A^{}$ using the counterterm, in agreement with the familiar result that $U(1)_V$ and $U(1)_A$ cannot be simultaneously conserved in the $VVA$ triangle diagram. Also, as discussed in \cref{footnote:eps_ct}, there is always a relevant $U(1)_A^3$ anomaly which cannot be removed by counterterms.

\subsection{Product of Abelian and Non-Abelian Sectors}
\label{subsec:U1Product}

Finally, we consider the cross terms in $\Acal_\beta^\Lambda[\alpha]$ between Abelian and non-Abelian sectors. These include the $\tr(\{t^a , t^b \}\, t^c)$ terms in \cref{eqn:Master2} with two of the adjoint indices in the same non-Abelian sector and the third index in an $U(1)$ sector, and the $\tr(t^at^bt^ct^d)$ terms with two of the adjoint indices in the same non-Abelian sector and the other two in either one or two $U(1)$ sectors. So in what follows we focus on a theory with one simple non-Abelian sector and up to two $U(1)$ sectors, which we call $U(1)_A$ and $U(1)_B$. To ease the presentation we reserve the notation $G^\mu$, $F^{\mu\nu}$, $\alpha$, $t^a$ that we have been using in the general calculation for the non-Abelian sector here, while denoting the corresponding objects in the $U(1)$ sectors by $A^\mu$, $F_A^{\mu\nu}$, $\alpha_A$, $Q_A$ and $B^\mu$, $F_B^{\mu\nu}$, $\alpha_B$, $Q_B$. We use $\bna$ to represent the common $\beta$ parameter associated with all the non-Abelian generators, and use $\beta_A$, $\beta_B$ for the $\beta$ parameters of the $U(1)$ sectors.

From the discussion in \cref{subsec:simple_nonAbelian} we know that the only values of $\bna$ consistent with the Wess-Zumino condition in the non-Abelian sector are 1 and 0. Let us first consider the simpler $\bna=1$ case. Here the $\tr(t^at^bQ_AQ_B)$ terms are all multiplied by $(1-\bna)$ and vanish, while for the $\tr(t^at^bQ_A)$ terms we have  (switching to matrix notation and following \cref{eqn:Master1}):
\begin{align}
\Acal_\beta^\Lambda [\alpha] &\supset \int \dd^4x \left( -\frac{1}{32\pi^2} \right) \eps_{\mu\nu\rho\sigma}
\tr \Big[ F^{\mu\nu} F^{\rho\sigma} \alpha_A 
\notag\\
&\hspace{150pt}
+ 2\beta_A F^{\mu\nu} F_A^{\rho\sigma} \alpha + 2(1-\beta_A) F^{\mu\nu}A^\rho (D^\sigma\alpha) \Big]
\notag\\[5pt]
&= \int \dd^4x\, \biggl(-\frac{1}{32\pi^2}\biggr)\, \varepsilon_{\mu\nu\rho\sigma}
\tr \Big[ F^{\mu\nu} F^{\rho\sigma} \alpha_A + (1+\beta_A) F^{\mu\nu} F_A^{\rho\sigma} \alpha \Big] \,.
\label{eq:A_beta_NAxA_1}
\end{align}
To arrive at the last equation we have integrated by parts and used the Bianchi identity $\eps^{\mu\nu\rho\sigma} (D_\sigma F_{\mu\nu}) = 0$. Performing the BRST transformation, we find
\begin{align}
\delta_\brst \Acal_\beta^\Lambda [\omega] &\supset \int \dd^4x\,
\frac{i}{32\pi^2}\, (1+\beta_A)\, \eps_{\mu\nu\rho\sigma} \tr \big( F^{\mu\nu} F_A^{\rho\sigma} \omega^2 \big) \,.
\end{align}
So for these cross terms in the anomaly to be consistent with the Wess-Zumino condition, we must have
\begin{equation}
\beta_A = -1 \qquad\text{or}\qquad \tr(t^a t^b Q_A)=0 \qquad\qquad\text{($\bna=1$ case)} \,.
\end{equation}
As a result, \cref{eq:A_beta_NAxA_1} either vanishes due to $\tr(t^a t^b Q_A)=0$, in which case there is no crossed anomaly, or only the $F^{\mu\nu} F^{\rho\sigma} \alpha_A$ term survives; the latter cannot be obtained as a local counterterm variation and is therefore a relevant anomaly. In fact, we have just recovered the non-Abelian generalization of the $U(1)_V\times U(1)_A$ example in the previous subsection (\cf\ \cref{eqn:U1VU1Abeta1m1}): swapping $U(1)_V$ for a non-anomalous non-Abelian sector (recall that $\bna=1$ requires $\tr(\{t^a , t^b \}\, t^c)=0$) leads to the same crossed anomaly with a chiral $U(1)$.

Next we consider the other option, $\bna=0$, for the non-Abelian sector. In this case, both the $\tr(t^a t^b Q_A)$ and $\tr(t^a t^b Q_A Q_B)$ terms can be nonzero. After some algebra we can organize the $\tr(t^a t^b Q_A)$ terms into the following form:
\begin{align}
\Acal_\beta^\Lambda [\alpha] &\supset \int \dd^4x \left( -\frac{1}{96\pi^2} \right) \eps_{\mu\nu\rho\sigma}
\tr \bigg[
F^{\mu\nu} F^{\rho\sigma}\alpha_A +i G^\mu G^\nu F^{\rho\sigma}\alpha_A + 2 G^\mu G^\nu G^\rho G^\sigma\alpha_A
\nonumber\\[5pt]&\hspace{100pt}
+\frac{3}{2} (1+\beta_A) F_\text{lin}^{\mu\nu} F_A^{\rho\sigma} \alpha - (1-\beta_A) F^{\mu\nu} A^\rho (\partial^\sigma \alpha) \bigg] \,.
\label{eq:A_beta_NAxA_0_ttQ}
\end{align}
Among the five terms, three (first, third and fourth) are actually BRST-invariant. Overall, we find \cref{eq:A_beta_NAxA_0_ttQ} has the following BRST transformation:
\begin{align}
\delta_\brst \Acal_\beta^\Lambda [\omega] &\supset \int \dd^4x
\left( -\frac{1}{96\pi^2} \right) \beta_A\, \eps_{\mu\nu\rho\sigma} \tr \Big[ F^{\mu\nu} (\partial^\rho\omega)(\partial^\sigma \omega_A) \Big] \,.
\end{align}
For this to vanish, we need
\begin{equation}
\beta_A = 0 \qquad\text{or}\qquad \tr(t^a t^b Q_A)=0 \qquad\qquad\text{($\bna=0$ case)} \,.
\end{equation}
So the crossed anomaly in \cref{eq:A_beta_NAxA_0_ttQ} either vanishes due to $\tr(t^a t^b Q_A)=0$ or is contained in the general $\beta=0$ formula \cref{eq:A_beta=0} as a relevant anomaly.

Meanwhile, for the $\tr(t^a t^b Q_A Q_B)$ terms, we find:
\begin{align}
\Acal_\beta^\Lambda [\alpha] &\supset \int \dd^4x \left(-\frac{1}{48\pi^2}\right) \tr \bigg\{
(1-\beta_A) \Big( \{ G^\mu, G^\nu \} \,A_\mu + G^\mu G_\mu A^\nu \Big) (\partial_\nu \alpha_B)
\notag\\[5pt]
&\hspace{110pt}
+ (1-\beta_B) \Big( \{ G^\mu, G^\nu \} \,B_\mu + G^\mu G_\mu B^\nu \Big) (\partial_\nu \alpha_A)
\notag\\[5pt]
&\hspace{35pt}
+ 2(1-\beta_A)(1-\beta_B) \Big[ \left( A^\mu B^\nu + A^\nu B^\mu \right) G_\mu +A^\mu B_\mu G^\nu \Big] (\partial_\nu \alpha) \bigg\} \,,
\label{eq:A_beta_NAxA_0_ttQQ}
\end{align}
which transforms under BRST as:
\begin{align}
\delta_\brst \Acal_\beta^\Lambda [\omega] &\supset \int \dd^4x\,
\frac{1}{48\pi^2}\, \tr \bigg\{
(\beta_A-\beta_B) \Big[ 2G^\mu G^\nu (\partial_\mu \omega_A) + G^\mu G_\mu (\partial^\nu \omega_A) \Big] (\partial_\nu \omega_B)
\notag\\[5pt]
&\hspace{-20pt}
-2(1-\beta_A) \beta_B \Big[ G^\mu (\partial^\nu \omega) A_\mu + G^\nu (\partial^\mu \omega) A_\mu
+ G^\mu (\partial_\mu \omega) A^\nu \Big] (\partial_\nu \omega_B)
\notag\\[5pt]
&\hspace{-20pt}
-2(1-\beta_B) \beta_A \Big[ G^\mu (\partial^\nu \omega) B_\mu + G^\nu (\partial^\mu \omega) B_\mu
+ G^\mu (\partial_\mu \omega) B^\nu \Big] (\partial_\nu \omega_A) \bigg\} \,.
\end{align}
For this to vanish, we need
\begin{equation}
\beta_A = \beta_B = (0 \;\,\text{or}\;\, 1) \qquad\text{or}\qquad \tr(t^a t^b Q_A Q_B) = 0 \qquad\qquad\text{($\bna=0$ case)} \,.
\end{equation}
So the crossed anomaly in \cref{eq:A_beta_NAxA_0_ttQQ} either vanishes due to $\tr(t^a t^b Q_A Q_B)=0$ or $\beta_A=\beta_B=1$, or is contained in the general $\beta=0$ formula \cref{eq:A_beta=0} as an irrelevant anomaly.

For both cases discussed above, $\bna=1$ and $\bna=0$, the relevant part of the crossed anomaly is proportional to $\tr(t^a t^b Q_A)$, so the anomaly cancellation condition is contained in the standard one, $\tr\,\bigl(\{t^a,t^b\}\,t^c\bigr) = 0$.

\section{Discussion and Future Directions}
\label{sec:Discussion}

In this paper, we introduced a novel regularization prescription to calculate anomalies for global and gauge symmetries using CDE. The calculation was performed in $d=4$ spacetime dimensions, thereby avoiding any of the subtleties that arise when computing anomalies using dimensional regularization. The master formula obtained in this framework integrates various known results regarding anomalies.

In a companion paper~\cite{Paper2}, we will extend the formalism developed here to incorporate the effects of higher dimensional operators into the anomaly calculation. This has an immediate application to the Standard Model Effective Field Theory (SMEFT). Recently, arguments that the SMEFT is not anomalous were provided in Refs.~\cite{Bonnefoy:2020tyv, Feruglio:2020kfq}. In Ref.~\cite{Paper2}, we will give an explicit proof using CDE that SMEFT is non-anomalous when including operators with general scalar, vector, and tensor couplings to fermion bilinears.

In future work, we would like to apply this formalism to compute the EFTs that emerge when integrating out fermions with chiral couplings (for example, integrating out the top quark in the Standard Model). This is well-known to produce an EFT with a Wess-Zumino-Witten term \cite{Wess:1971yu, DHoker:1984izu,DHoker:1984mif}.  It should be possible to extend the calculations presented here to reproduce this result in a new way.  This will require understanding the interplay of the method presented here and the results for other functional traces that are evaluated using dimensional regularization, since the functional EFT matching framework relies on the method of regions, which is implemented in dimensional regularization.  At least for one loop calculations, the use of different regulators may not cause any particular difficulties.  Once this is understood, functional methods for one-loop matching will be a complete framework for integrating out any heavy particles with spins 0, 1/2, and 1.

%%%%%%%%%%%%%%%%%%%%%%%%%%%%%%%%%%%%%%%%%%%%%%%%%%%%%%%%%%%%%%%%%%%%%%%%%%%%%%%%
\acknowledgments
\addcontentsline{toc}{section}{\protect\numberline{}Acknowledgments}
%%%%%%%%%%%%%%%%%%%%%%%%%%%%%%%%%%%%%%%%%%%%%%%%%%%%%%%%%%%%%%%%%%%%%%%%%%%%%%%%

We thank Quentin Bonnefoy, Nathaniel Craig, Sungwoo Hong, Markus Luty and Aneesh Manohar for useful discussions. T.C.\ is supported by the U.S.\ Department of Energy under grant number DE-SC0011640. X.L.\ is supported by the U.S.\ Department of Energy under grant numbers DE-SC0009919 and DE-SC0011640. Z.Z.\ is supported by the U.S.\ Department of Energy under grant number DE-SC0011702. This work was performed in part at Aspen Center for Physics, which is supported by National Science Foundation grant PHY-1607611.

%%%%%%%%%%%%%%%%%%%%%%%%%%%%%%%%%%%%%%%%%%%%%%%%%%%%%%%%%%%%%%%%%%%%%%%%%%%%%%%%
\appendix
\section*{Appendix}
\addcontentsline{toc}{section}{\protect\numberline{}Appendix}
\renewcommand*{\thesubsection}{\Alph{subsection}}
\numberwithin{equation}{subsection}
%%%%%%%%%%%%%%%%%%%%%%%%%%%%%%%%%%%%%%%%%%%%%%%%%%%%%%%%%%%%%%%%%%%%%%%%%%%%%%%%

%%%%%%%%%%%%%%%%%%%%%%%%%%%%%%%%%%%%%%%%%%%%%%%%%%%%%%%%%%%%%%%%%%%%%%%%%%%%%%%%
\subsection{Comments on Cyclic Permutation}
\label{appsec:perm}
%%%%%%%%%%%%%%%%%%%%%%%%%%%%%%%%%%%%%%%%%%%%%%%%%%%%%%%%%%%%%%%%%%%%%%%%%%%%%%%%

In this appendix, we clarify a subtle point in performing CDE calculations, \ie, when (and why) we are allowed to perform cyclic permutations on the argument of a functional trace `$\,\Tr\left(\cdots\right)\,$', and a lowercase trace `$\,\tr\left(\cdots\right)\,$' which is only over the internal indices.

We begin by recalling that a functional operator $\O$ is a matrix that acts on both the functional vector space $\ket{x}$ and some internal vector space. The latter is typically finite dimensional, which we can label by a discrete index $i$. We can then write out the concrete relation between the functional trace `$\,\Tr\,$' and the internal trace `$\,\tr\,$':
\begin{equation}
\Tr \left(\O\right) = \int \dd^4x\, \bra{x} \tr\left(\O\right) \ket{x} = \int \dd^4x\, \bra{x} \O_{ii} \ket{x} \,.
\label{eqn:Trtrdef}
\end{equation}
Clearly, the functional trace `$\Tr$' sums over all the indices of the matrix $\O$, and therefore it is always safe to perform a cyclic permutation:
\begin{align}
\Tr \left( \O^A \O^B \right) &= \int \dd^4x\, \dd^4y\, \bra{x} \O^A_{ij} \ket{y} \bra{y} \O^B_{ji} \ket{x}
\notag\\[5pt]
&= \int \dd^4y\, \dd^4x\, \bra{y} \O^B_{ji} \ket{x} \bra{x} \O^A_{ij} \ket{y}
= \Tr \left( \O^B \O^A \right) \,.
\label{eqn:TrPermutation}
\end{align}
On the other hand, the internal trace `$\tr$' only sums over a subset of indices for the matrix $\O$, and therefore it is generically illegal to make cyclic permutations inside `$\tr$' alone:
\begin{equation}
\tr \left( \O^A \O^B \right) \ne \tr \left( \O^B \O^A \right)
\quad\Longleftrightarrow\quad
\bra{x} \tr \left( \O^A \O^B \right) \ket{y} \ne \bra{x} \tr \left( \O^B \O^A \right) \ket{y} \,.
\label{eqn:trNoPermutation}
\end{equation}
Note that after taking the internal trace, the object $\tr \left( \O^A \O^B \right)$ is still a matrix acting on the functional space spanned by $\ket{x}$. So when we check whether the two objects $\tr \left( \O^A \O^B \right)$ and $\tr \left( \O^B \O^A \right)$ are equal, it is a comparison of two matrices where one needs to compare entry by entry, as indicated by the right-hand expression of \cref{eqn:trNoPermutation}. Generically, they are not equal and making cyclic permutations inside `$\tr$' alone is not allowed.

However, in many practical calculations of functional traces, the evaluation results (after carrying out the loop integrals) are \emph{local} action-like expressions that generically have the form (see \eg~\cref{eqn:Master0,eqn:DiracTraces})
\begin{equation}
\Tr \left( \cdots \right) = \int \dd^4x\; \ttr \left( \O^A \O^B \O^C \cdots \right) \,,
\label{eqn:ttrInResults}
\end{equation}
where the reason for using a slightly different notation `$\,\ttr\left(\cdots\right)\,$' will become clear shortly. When handling expressions like \cref{eqn:ttrInResults}, we do sometimes make cyclic permutations to simplify the calculation:
\begin{equation}
\text{Sometimes we take}:\qquad
\int \dd^4x\; \ttr \left( \O^A \O^B \right) = \int \dd^4x\; \ttr \left( \O^B \O^A \right) \,.
\label{eqn:ttrPermutation}
\end{equation}
This has been used extensively in \cref{sec:Evaluation}, as well as for many functional matching calculations with CDE in the literature. The purpose of this appendix is to clarify when and why \cref{eqn:ttrPermutation} could hold. The explanation has two important aspects:
\begin{itemize}
\item There is a slight abuse of notation `$\,\tr\,$' in expressions like \cref{eqn:Master0,eqn:DiracTraces}. As emphasized by using a different notation `$\,\ttr\,$' above, the traces in \cref{eqn:ttrInResults,eqn:ttrPermutation} are not precisely the same objects as the traces in \cref{eqn:trNoPermutation} -- the latter are matrices acting on the functional space $\ket{x}$, while the former are actually elements of those matrices.
\item \cref{eqn:ttrPermutation} does not hold for generic operators $\O^A$, $\O^B$. However, if both $\O^A\O^B$ and $\O^B\O^A$ are diagonal functional operators in the position basis $\ket{x}$, namely if they satisfy
\begin{subequations}\label{eqn:ttrCondition}
\begin{align}
\tr \left(\O^A \O^B\right) \ket{x} &= t^{AB} (x) \ket{x} \,, \\[5pt]
\tr \left(\O^B \O^A\right) \ket{x} &= t^{BA} (x) \ket{x} \,.
\end{align}
\end{subequations}
for some ordinary functions $t^{AB}(x)$ and $t^{BA}(x)$, then \cref{eqn:ttrPermutation} holds.
\end{itemize}
In what follows, we elaborate on these two aspects in turn.

\subsubsection{Internal Trace Notation}
\label{appsubsec:trNotations}

First, it is clear from \cref{eqn:ttrInResults} that $\ttr \left( \O \right)$ must be an ordinary function of the variable $x$ (similar to a Lagrangian), such that the integral in \cref{eqn:ttrInResults} would yield a local action-like result. So $\ttr \left( \O \right)$ cannot be a matrix on the functional space $\ket{x}$. Instead, it should be interpreted as an element of that matrix.

Second, we emphasize that $\ttr \left( \O \right)$ is not the following matrix element that one might naively expect:
\begin{equation}
\ttr \left( \O \right) \ne \bra{x} \tr \left( \O \right) \ket{x} \,.
\end{equation}
If the above were true, then performing the integral in \cref{eqn:ttrPermutation} would give us the functional trace
\begin{equation}
\int \dd^4x\, \bra{x} \tr \left( \O^A \O^B \right) \ket{x} = \Tr \left( \O^A \O^B \right) \,,
\end{equation}
in which cyclic permutation would not be a problem at all, as explained around \cref{eqn:TrPermutation}. But it is clear that \cref{eqn:ttrPermutation} is not supposed to yield $\Tr \left( \O^A \O^B \right)$. The correct matrix element is
\begin{equation}
\ttr \left( \O \right) = \int \dd^4y\, \bra{x} \tr \left( \O \right) \ket{y} \,.
\label{eqn:ttrMeaning}
\end{equation}
To understand this subtle point, we need to remind ourselves how we usually obtain expressions like \cref{eqn:ttrInResults} and hence terms like $\ttr \left( \O \right)$ from the CDE evaluation. Usually, we start with a functional trace like \cref{eqn:TbetaLambdak} and calculate it using momentum eigenstates:
\begin{equation}
\Tr \left[ f\left( i\hat\partial_\mu, U(\hat{x}) \right) \right]
= \int \frac{\dd^4q}{(2\pi)^4}\, \mel**{q}{\,\tr \left[ f\left( i\hat\partial_\mu, U(\hat{x}) \right) \right] }{q}  \,.
\label{eqn:Trfdef}
\end{equation}
Using the fact
\begin{equation}
\ket{q} = \int \dd^4x\, \ket{x}\bra{x}\ket{q} = \int \dd^4x\, e^{-iqx} \ket{x} = \int \dd^4x\, e^{-iq\hat{x}} \ket{x} \,,
\end{equation}
we can rewrite \cref{eqn:Trfdef} as
\begin{align}
\Tr \left[ f\left( i\hat\partial_\mu, U(\hat{x}) \right) \right]
&= \int \dd^4x\, \dd^4y\, \int \frac{\dd^4q}{(2\pi)^4}\,
\mel**{x}{e^{iq\hat{x}} \,\tr \left[ f\left( i\hat\partial_\mu, U(\hat{x}) \right) \right] e^{-iq\hat{x}}}{y}
\notag\\[5pt]
&= \int \dd^4x\, \dd^4y\, \int \frac{\dd^4q}{(2\pi)^4}\,
\mel**{x}{ \,\tr \left[ f\left( q_\mu + i\hat\partial_\mu, U(\hat{x}) \right) \right] }{y}
\notag\\[5pt]
&= \int \dd^4x\, \left\{ \int \dd^4y\,
\mel**{x}{ \int \frac{\dd^4q}{(2\pi)^4}\, \tr \left[ f\left( q_\mu + i\hat\partial_\mu, U(\hat{x}) \right) \right] }{y} \right\}
\notag\\[5pt]
&= \int \dd^4x\, \bigg\{ \int \dd^4y\,
\mel**{x}{ \,\tr \left( \O_f \right) }{y} \bigg\} \,,
\label{eqn:TrfCDE}
\end{align}
where $\O_f$ is defined implicitly by the last equation. As indicated in the last line, one way of understanding the `simplified CDE' is that one Taylor expands the function `$\,f\,$' above and performs the momentum loop integral over $q_\mu$ to obtain a set of functional operators of the form $\tr \left( \O_f \right)$. This is precisely what we did in deriving \cref{eqn:Master0,eqn:DiracTraces} from \cref{eqn:TbetaLambdak}. Now comparing \cref{eqn:TrfCDE} with \cref{eqn:ttrInResults}, we see that the notation `$\,\ttr\,$' is actually denoting the quantity inside the curly brackets in \cref{eqn:TrfCDE}. Therefore, we have carefully derived the relation in \cref{eqn:ttrMeaning}.

Let us recall that the definition of the `functional vector space' is the collection of all the functions $\phi(x)$ (usually satisfying certain constraints, such as $\norm{\phi}_2<\infty$ (under box normalization)), where each function corresponds to a vector $\ket{\phi}$:
\begin{equation}
\phi(x) = \bra{x}\ket{\phi} \,.
\end{equation}
It thus provides us with a linear algebra language for the differential operations. Specifically, the process of a differential operator $\hat{f}$ acting on a function $\phi(x)$ to yield a new function $\bigl( \hat{f} \phi \bigr) (x)$ can be written as the action of a matrix in this linear space:
\begin{equation}
\bigl( \hat{f} \phi \bigr) (x) = 
\bigl\langle x \big| \hat{f}\phi\bigr\rangle
= \bigl\langle x \big| \hat{f} \big| \phi\bigr\rangle
= \int \dd^4y\, 
\bigl\langle x \big| \hat{f} \big| y\bigr\rangle
\bra{y}\ket{\phi} \,.
\label{eqn:Differential}
\end{equation}
The key to this dictionary are the matrix elements $\bigl\langle x \big| \hat{f} \big| y\bigr\rangle$ for various differential operators. When $\hat{f}$ is an ordinary function such as $\hat{f} = G_\mu(x)$, its matrix is diagonal in the $\ket{x}$ basis:
\begin{subequations}
\begin{align}
\mel**{x}{G_\mu}{y} &= G_\mu (x)\, \delta^4(x-y) \,, \\[5pt]
G_\mu(x) \phi(x) &= \int \dd^4y\, \mel**{x}{G_\mu}{y} \bra{y}\ket{\phi} = \int \dd^4y \left[ G_\mu (x)\, \delta^4(x-y) \right] \phi(y) \,.
\end{align}
\end{subequations}
When $\hat{f} = \partial_\mu$ is a derivative, we have 
\begin{subequations}
\begin{align}
\mel**{x}{\partial_\mu}{y} &= \frac{\partial}{\partial x^\mu}\, \delta^4(x-y) \,, \label{eqn:pdelement} \\[5pt]
\partial_\mu \phi (x) &= \int \dd^4y\, \mel**{x}{\partial_\mu}{y} \bra{y}\ket{\phi} = 
\frac{\partial}{\partial x^\mu}\,\int \dd^4y\, \delta^4(x-y) \,\phi(y) \,.
\end{align}
\end{subequations}
General differential operators, like $\hat{f}\big( i\hat\partial_\mu, U(\hat{x}) \big)$ in \cref{eqn:TrfCDE}, are built from the two kinds of operators discussed above.

We note in particular that the constant unity function `$1$' corresponds to a vector $\ket{1}$ that satisfies
\begin{equation}
\ket{1} = \int \dd^4y\, \ket{y}\bra{y}\ket{1} = \int \dd^4y\, \ket{y} \,.
\end{equation}
Therefore, the relation in \cref{eqn:ttrMeaning} can be rewritten as
\begin{equation}
\ttr \left( \O \right) = \bra{x} \tr \left( \O \right) \ket{1} = \Big( \tr \left( \O \right) 1 \Big) (x) \,,
\label{eqn:ttrDifferential}
\end{equation}
where the last expression follows from the differential operation language in \cref{eqn:Differential} -- we are simply taking the differential operator $\tr \left( \O \right)$, acting it on the constant unity function $1$, and then evaluating the resulting function at point $x$.\footnote{See \eg\ Sec.~2.2 of Ref.~\cite{Henning:2014wua} and App.~B.2.2 of Ref.~\cite{Cohen:2019btp} for clarifications of this point.}
When the function being acted on is the constant unity function $1$, we often suppress it. We also often suppress the explicit `$(x)$' when talking about a function. Doing both for the last expression in \cref{eqn:ttrDifferential} leads to our abuse of the notation `$\tr$' in the main text.

From \cref{eqn:ttrDifferential}, it is immediately clear that
\begin{subequations}\label{eqn:zeros}
\begin{align}
\ttr \left( A B \cdots C \partial_\mu \right) &= 0 \,, \\[5pt]
\ttr \left( \partial_\mu A B \cdots C \right) &\text{ is a total derivative.}
\end{align}
\end{subequations}
With these, we can see a quick counterexample to \cref{eqn:ttrPermutation}:
\begin{subequations}\label{eqn:counterexample}
\begin{align}
\ttr \left( A \partial_\mu B^\mu \right) &= \ttr \big[ A \left( \partial_\mu B^\mu \right) + A B^\mu \partial_\mu \big] 
= \ttr \big[ A \left(\partial_\mu B^\mu \right) \big] \,, \label{eqn:pdClosedonB} \\[5pt]
\ttr \left( B^\mu A \partial_\mu \right) &= 0 \,.
\end{align}
\end{subequations}
Clearly, the two lines are related by a cyclic permutation of $B^\mu$, but they are generically not equal.

\subsubsection{Conditions for Cyclic Permutations in Internal Traces}
\label{appsubsec:Condition}

After clarifying the meaning of `$\,\ttr\left(\cdots\right)\,$', namely the relation in \cref{eqn:ttrMeaning}, we see that \cref{eqn:ttrPermutation} does not always hold. However, if both expressions inside the trace before and after the cyclic permutation are diagonal operators in the position basis $\ket{x}$, \ie, if \cref{eqn:ttrCondition} is true, then \cref{eqn:ttrPermutation} would hold.

To see this, we first note that if 
\begin{equation}
\tr \left( \O^A \O^B \right) \ket{x} = t^{AB} (x) \ket{x} \,,
\end{equation}
then we simply have
\begin{equation}
\ttr \left( \O^A \O^B \right) = \int \dd^4y\, \mel**{x}{\,\tr \left(\O^A \O^B\right)}{y} = \int \dd^4y\, t^{AB}(y)\, \delta^4(x-y) = t^{AB}(x) \,.
\end{equation}
Therefore, it is linked with the functional trace as
\begin{align}
\Tr \left( \O^A \O^B \right) &= \int \dd^4x \int \frac{\dd^4q}{(2\pi)^4}\, \bra{x}\ket{q} \mel**{q}{\,\tr \left( \O^A \O^B \right) }{x} \notag\\[5pt]
&= \int \dd^4x\, t^{AB}(x) \int \frac{\dd^4q}{(2\pi)^4}\bra{x}\ket{q}\bra{q}\ket{x} 
= \int \dd^4x\; \ttr \left( \O^A \O^B \right) \int \frac{\dd^4q}{(2\pi)^4} \,.\notag\\[5pt]
\end{align}
where $\int \frac{\dd^4q}{(2\pi)^4} = \bra{x}\ket{x}$ is just a normalization factor. Making use of this relation between $\ttr(\cdots)$ and $\Tr(\cdots)$, one could take advantage of \cref{eqn:TrPermutation} to perform a cyclic permutation:
\begin{align}
\int \dd^4x\; \ttr \left( \O^A \O^B \right) \int \frac{\dd^4q}{(2\pi)^4} &= \Tr \left( \O^A \O^B \right) \notag\\[5pt]
&= \Tr \left( \O^B \O^A \right) = \int \dd^4x\; \ttr \left( \O^B \O^A \right) \int \frac{\dd^4q}{(2\pi)^4} \,.
\label{eqn:ttrViaTr}
\end{align}
Canceling the normalization factor gives us \cref{eqn:ttrPermutation}.

Note that one can generalize \cref{eqn:ttrPermutation} to the sum of multiple terms:
\begin{equation}
\O^A \O^B \quad\longrightarrow\quad
\O_1^A \O_1^B + \cdots + \O_n^A \O_n^B \,.
\end{equation}
In this case, for the steps in \cref{eqn:ttrViaTr} to be valid, one only needs the sum to be diagonal in the position basis $\ket{x}$, namely we have
\begin{equation}
\hspace{-10pt}
\int \dd^4x\; \ttr \left( \O_1^A \O_1^B + \cdots + \O_n^A \O_n^B \right)
= \int \dd^4x\; \ttr \left( \O_1^B \O_1^A + \cdots + \O_n^B \O_n^A \right) \,,
\end{equation}
provided that
\begin{subequations}\label{eqn:ttrConditionMultiple}
\begin{align}
\tr \left( \O_1^A \O_1^B + \cdots + \O_n^A \O_n^B \right) \ket{x} &= t^{AB} (x) \ket{x} \,, \\[5pt]
\tr \left( \O_1^B \O_1^A + \cdots + \O_n^B \O_n^A \right) \ket{x} &= t^{BA} (x) \ket{x} \,.
\end{align}
\end{subequations}

An operator $\O$ being diagonal in the position basis $\ket{x}$ is equivalent to the statement that all the derivatives in $\O$ are \emph{closed}. For example, consider the following differential operator:
\begin{align}
\O = A \partial_\mu B C &= A \big( \partial_\mu B \big) C + A B \partial_\mu C \notag\\[5pt]
&= A \big( \partial_\mu B \big) C + A B \big( \partial_\mu C \big) + A B C \partial_\mu \,,
\label{eqn:openvsclosed}
\end{align}
where $A, B, C$ are diagonal in the $\ket{x}$ basis. The decomposition in the first line follows from the product rule of the derivative, where the parentheses in the first term has the usual interpretation -- it indicates that $\partial_\mu$ only acts on $B$ but not anything to the right of $B$. (In fact, this notation was already used in \cref{eqn:pdClosedonB}.) In this case, we say that the derivative is \emph{closed} on $B$. In contrast, the second term in the first line has an \emph{open} derivative that acts on everything to its right. One can further use the product rule to obtain the decomposition in the second line, where a term with the derivative closed on $C$ appears, and there is an additional term with an open derivative. Clearly, terms with closed derivatives, such as $A \big( \partial_\mu B \big) C$ and $A B \big( \partial_\mu C \big) $ are diagonal operators in the $\ket{x}$ basis, while terms with open derivatives such as $A B C \partial_\mu$ are not; see \eg\ \cref{eqn:pdelement}.

When evaluating a functional trace with simplified CDE, the initial set of operators in the trace $\ttr (\cdots)$ emerge from evaluating an expression of the form (see \cref{eqn:TrfCDE}):
\begin{equation}
\int \frac{\dd^4q}{(2\pi)^4}\, \tr \left[ f\left( q_\mu + i\hat\partial_\mu, U(\hat{x}) \right) \right] = \tr \left( \O_f \right) \,.
\end{equation}
The operator $\tr \left( \O_f \right)$ derived from such an expression, \ie, upon expanding `$\,f\,$' and carrying out the loop momentum integral, is guaranteed to be diagonal in the position basis $\ket{x}$, because it is known that one could use the trick of `original CDE'  to close all of the derivatives in it (see \eg\ App.~B.2.3 of Ref.~\cite{Cohen:2019btp}). However, since $\O_f$ is a sum of terms, if we perform an arbitrary cyclic permutation on each term:
\begin{equation}
\tr \left( \O_f \right) = \tr \left( \O_1^A \O_1^B + \cdots + \O_n^A \O_n^B \right) 
\quad\longrightarrow\quad
\tr \left( \O_1^B \O_1^A + \cdots + \O_n^B \O_n^A \right) \,,
\end{equation}
it is not guaranteed that the operator is still diagonal in the $\ket{x}$ basis, thus invalidating the operation. Only a subset of cyclic permutations that satisfy the condition in \cref{eqn:ttrConditionMultiple} are `legal.'

Nonetheless, in practical calculations, a very efficient prescription to ensure that we are only performing legal cyclic permutations is to \emph{stipulate} that terms with \emph{open} derivatives should not be evaluated -- one must keep track of all such terms, and make sure that they get canceled upon summing the terms obtained after the cyclic permutations. If they do not get fully canceled, then it is a sign that an illegal cyclic permutation had been carried out. In this case, one needs to make further cyclic permutations until the derivatives are all closed. In summary, insisting that all derivatives must be closed in the end is an efficient way to make sure that we are carrying out legal cyclic permutations. The calculations in \cref{sec:Evaluation} of the main text (as well as in many other functional matching calculations with CDE in the literature) are done in such a manner.

Let us look at a quick example of this:
\begin{align}
\ttr \big[ \left( \partial_\mu A \right) B^\mu \big] &\;\,=\;\, \ttr \big[ \partial_\mu A B^\mu - A \partial_\mu B^\mu \big] \notag\\[5pt]
&\longrightarrow \ttr \big[ \partial_\mu A B^\mu - B^\mu A \partial_\mu \big] =  \ttr \big[ (\partial_\mu A B^\mu) + A B^\mu \partial_\mu - B^\mu A \partial_\mu \big] \notag\\[5pt]
&\longrightarrow \ttr \big[ B^\mu \partial_\mu A - B^\mu A \partial_\mu \big] = \ttr \big[ B^\mu \left( \partial_\mu A \right) \big] \,.
\label{eqn:cycexample}
\end{align}
In the first line, we started with an operator $\left( \partial_\mu A \right) B^\mu$ in which the derivative is closed. We made a cyclic permutation of the second term to arrive at the second line, where the derivatives are not fully closed, because the last two terms in the second expression both have open derivatives and they do not cancel each other. If we were to stop here and evaluate the second line, then following \cref{eqn:zeros} these terms are zero and total derivatives that would not feed into the final result:
\begin{equation}
\int \dd^4x\; \ttr \big[ (\partial_\mu A B^\mu) + A B^\mu \partial_\mu - B^\mu A \partial_\mu \big] = 0 \,.
\end{equation}
This clearly would not agree with the evaluation of the left-hand side of the first line. The reason is that the second line was obtained by an illegal cyclic permutation. Now, if we insist that the second line of \cref{eqn:cycexample} should not be evaluated since it has open derivatives, then we are forced to make further cyclic permutations such that all the derivatives can be closed upon summing the terms. The third line is an example of such a further cyclic permutation. As soon as the derivatives are all closed, we can carry out the evaluation. This prescription guarantees that only legal cyclic permutations would be performed, and we can see that the result obtained in the third line does agree with the expression we started with in the first line.

\end{spacing}

\begin{spacing}{1.09}
\addcontentsline{toc}{section}{\protect\numberline{}References}
\bibliographystyle{utphys}
\bibliography{Anomalies}
\end{spacing}

\end{document}